\documentclass{aastex631}

\def\sgr {SGR 1935+2154}
\usepackage{threeparttable}
 \usepackage{amsmath}
\usepackage{natbib}
\usepackage{sansmath}
\usepackage{multirow}

\usepackage{xcolor}
\usepackage{array}

\begin{document}
\title{X-ray hardening preceding the onset of SGR 1935+2154's radio pulsar phase}

\correspondingauthor{Di Li, Jian Li}
\email{dili@nao.cas.cn, jianli@ustc.edu.cn}

\author[0000-0002-3386-7159]{Pei Wang}
\affiliation{National Astronomical Observatories, Chinese Academy of Sciences, Beijing 100101, China}
\affiliation{Institute for Frontiers in Astronomy and Astrophysics, Beijing Normal University, Beijing 102206, China}

\author[0000-0003-1720-9727]{Jian Li}
\affiliation{CAS Key Laboratory for Research in Galaxies and Cosmology, Department of Astronomy, University of Science and Technology of China, Hefei 230026, China}
\affiliation{School of Astronomy and Space Science, University of Science and Technology of China, Hefei 230026, China}

\author[0000-0001-9599-7285]{Long Ji}
\affiliation{School of Physics and Astronomy, Sun Yat-Sen University, Zhuhai, 519082, China}

\author[0000-0003-0933-6101]{Xian Hou}
\affiliation{Yunnan Observatories, Chinese Academy of Sciences, Kunming 650216, China}
\affiliation{Key Laboratory for the Structure and Evolution of Celestial Objects, Chinese Academy of Sciences, Kunming 650216, China}

\author{Erbil G\"{u}gercino\u{g}lu}
\affiliation{National Astronomical Observatories, Chinese Academy of Sciences, Beijing 100101, China}
\affiliation{Istanbul University, Faculty of Science, Department of Astronomy and Space Sciences, Beyaz{\i}t, 34119, Istanbul, Turkey}

\author[0000-0003-3010-7661]{Di Li}
\affiliation{Department of Astronomy, Tsinghua University, Beijing 100084, China}
\affiliation{National Astronomical Observatories, Chinese Academy of Sciences, Beijing 100101, China}

\author[0000-0002-1522-9065]{Diego F. Torres}
\affiliation{Institute of Space Sciences (ICE, CSIC), Campus UAB, Carrer de Can Magrans, 08193, Barcelona, Spain}
\affiliation{Institut d'Estudis Espacials de Catalunya (IEEC), 08860 Castelldefels (Barcelona), Spain}
\affiliation{Instituci\'o Catalana de Recerca i Estudis Avan\c{c}ats (ICREA), E-08010, Barcelona, Spain}

\author{Yutong Chen}
\affiliation{National Astronomical Observatories, Chinese Academy of Sciences, Beijing 100101, China}
\affiliation{University of Chinese Academy of Sciences, Beijing 100049, China}

\author[0000-0001-8065-4191]{Jiarui Niu}
\affiliation{National Astronomical Observatories, Chinese Academy of Sciences, Beijing 100101, China}
\affiliation{University of Chinese Academy of Sciences, Beijing 100049, China}

\author[0000-0001-5105-4058]{Wei-Wei Zhu}
\affiliation{National Astronomical Observatories, Chinese Academy of Sciences, Beijing 100101, China}
\affiliation{Institute for Frontiers in Astronomy and Astrophysics, Beijing Normal University, Beijing 102206, China}

\author[0000-0002-9725-2524]{Bing Zhang}
\affiliation{Department of Physics and Astronomy, University of Nevada, Las Vegas, Las Vegas, NV 89154, USA}

\author[0000-0002-7044-733X]{En-Wei Liang}
\affiliation{Guangxi Key Laboratory for Relativistic Astrophysics, School of Physical Science and Technology, Guangxi University, Nanning 530004, China}

\author{Li Zhang}
\affiliation{Department of Astronomy, Key Laboratory of Astroparticle Physics of Yunnan Province, Yunnan University, Kunming 650091, China}

\author[0000-0002-3776-4536]{Mingyu Ge}
\affiliation{Key Laboratory of Particle Astrophysics, Institute of High Energy Physics, Chinese Academy of Sciences, Beijing 100049, China}

\author[0000-0002-7835-8585]{Zigao Dai}
\affiliation{CAS Key Laboratory for Research in Galaxies and Cosmology, Department of Astronomy, University of Science and Technology of China, Hefei 230026, China}
\affiliation{School of Astronomy and Space Science, University of Science and Technology of China, Hefei 230026, China}

\author{Lin Lin}
\affiliation{Institute for Frontiers in Astronomy and Astrophysics, Beijing Normal University, Beijing 102206, China}

\author[0000-0002-9274-3092]{Jinlin Han}
\affiliation{National Astronomical Observatories, Chinese Academy of Sciences, Beijing 100101, China}

\author{Yi Feng}
\affiliation{Zhejiang Lab, Hangzhou, Zhejiang 311121, China}

\author{Chenhui Niu}
\affiliation{Institute of Astrophysics, Central China Normal University, Wuhan 430079, China}

\author[0000-0002-8744-3546]{Yongkun Zhang}
\affiliation{National Astronomical Observatories, Chinese Academy of Sciences, Beijing 100101, China}
\affiliation{University of Chinese Academy of Sciences, Beijing 100049, China}

\author[0000-0002-6423-6106]{Dejiang Zhou}
\affiliation{National Astronomical Observatories, Chinese Academy of Sciences, Beijing 100101, China}
\affiliation{University of Chinese Academy of Sciences, Beijing 100049, China}

\author{Heng Xu}
\affiliation{National Astronomical Observatories, Chinese Academy of Sciences, Beijing 100101, China}
\affiliation{Department of Astronomy, Peking University, Beijing 100871, China}

\author[0000-0002-4327-711X]{Chunfeng Zhang}
\affiliation{National Astronomical Observatories, Chinese Academy of Sciences, Beijing 100101, China}
\affiliation{Department of Astronomy, Peking University, Beijing 100871, China}

\author[0000-0002-6465-0091]{Jinchen Jiang}
\affiliation{National Astronomical Observatories, Chinese Academy of Sciences, Beijing 100101, China}
\affiliation{Department of Astronomy, Peking University, Beijing 100871, China}

\author[0000-0002-9441-2190]{Chenchen Miao}
\affiliation{Zhejiang Lab, Hangzhou, Zhejiang 311121, China}

\author[0000-0003-1874-0800]{Mao Yuan}
\affiliation{National Space Science Center, Chinese Academy of Sciences, Beijing 101400, China}

\author{Weiyang Wang}
\affiliation{University of Chinese Academy of Sciences, Beijing 100049, China}
\affiliation{Kavli institute for astronomy and astrophysics, Peking University; Beijing 100871, China}

\author[0000-0002-7420-9988]{Dengke Zhou}
\affiliation{Zhejiang Lab, Hangzhou, Zhejiang 311121, China}

\author[0000-0001-9956-6298]{Jianhua Fang}
\affiliation{Zhejiang Lab, Hangzhou, Zhejiang 311121, China}

\author[0000-0003-4415-2148]{Youling Yue}
\affiliation{National Astronomical Observatories, Chinese Academy of Sciences, Beijing 100101, China}

\author{Yunsheng Wu}
\affiliation{Tencent Youtu Lab, Shanghai 201103, China}

\author[0000-0002-6592-8411]{Yabiao Wang}
\affiliation{Tencent Youtu Lab, Shanghai 201103, China}

\author{Chengjie Wang}
\affiliation{Tencent Youtu Lab, Shanghai 201103, China}

\author{Zhenye Gan}
\affiliation{Tencent Youtu Lab, Shanghai 201103, China}

\author{Yuxi Li}
\affiliation{Tencent Youtu Lab, Shanghai 201103, China}

\author{Zhongyi Sun}
\affiliation{Tencent Youtu Lab, Shanghai 201103, China}

\author{Mingmin Chi}
\affiliation{Fudan University, Shanghai 200438, China}

\author[0009-0005-8586-3001]{Junshuo Zhang}
\affiliation{National Astronomical Observatories, Chinese Academy of Sciences, Beijing 100101, China}
\affiliation{University of Chinese Academy of Sciences, Beijing 100049, China}

\author[0009-0000-7501-2215]{Jinhuang Cao}
\affiliation{National Astronomical Observatories, Chinese Academy of Sciences, Beijing 100101, China}
\affiliation{University of Chinese Academy of Sciences, Beijing 100049, China}

\author[0000-0001-5653-3787]{Wanjin Lu}
\affiliation{National Astronomical Observatories, Chinese Academy of Sciences, Beijing 100101, China}
\affiliation{University of Chinese Academy of Sciences, Beijing 100049, China}

\author[0000-0002-7372-4160]{Yidan Wang}
\affiliation{National Astronomical Observatories, Chinese Academy of Sciences, Beijing 100101, China}
\affiliation{University of Chinese Academy of Sciences, Beijing 100049, China}

\begin{abstract}
Magnetars are neutron stars with extremely strong magnetic fields, frequently powering high-energy 
activity in X-rays.
Pulsed radio emission following some X-ray outbursts have been detected (\citealt{Camilo2006,camilo2007a}), 
albeit its physical origin is unclear.
It has long been speculated that the origin of magnetars' radio signals is different from those from canonical pulsars, although convincing evidence is still lacking.
Five months after magnetar SGR 1935+2154's X-ray outburst and its associated Fast Radio Burst (FRB) 20200428, a radio pulsar phase was discovered.
Here we report the discovery of X-ray spectral hardening associated with the emergence of periodic radio pulsations from SGR 1935+2154 and a detailed analysis of the properties of the radio pulses. 
The observations suggest that radio emission originates from the outer magnetosphere of the magnetar, and the surface heating due to the bombardment of inward-going particles from the radio emission region is responsible for the observed X-ray spectral hardening. 
\end{abstract}

\section{Introduction} \label{sec:intro}
\sgr\/ is a Galactic magnetar discovered in 2014 by Swift-BAT (\citealt{2014GCN.16520....1S}). Its spin period ($\sim\/$3.24 s) and spin-down rate imply a surface magnetic field of $\sim2.2\times10^{14}$ G (\citealt{2016MNRAS.457.3448I}). Its spatial consistency with a supernova remnant (SNR) G57.2+0.8 suggests an actual association, leading to a distance estimation of 6.6$\pm$0.7 kpc (\citealt{zhou2020}). Since its discovery, \sgr\/ has shown multiple outbursts: in 2015 February, 2016 May, 2019 November, 2020 April, 2021 September and 2022 October (\citealt{Kozlova2016, 2017ApJ...847...85Y,2021outburst, 2022outburst}). Simultaneous with a hard X-ray burst (\citealt{2021NatAs...5..378L, Mereghetti2020, 2021NatAs...5..372R,2021NatAs...5..401T}) on 28th April 2020, \sgr\/ was observed to be associated with the first Galactic Fast Radio Burst (FRB) FRB 20200428 (\citealt{2020Natur.587...54C,2020Natur.587...59B}), thus providing the first confirmed link between FRBs and magnetars  (\citealt{2019A&ARv..27....4P, 2019ARA&A..57..417C}). This discovery promoted intense follow up observations with a number of radio telescopes. However, only a few radio bursts were detected and the radio pulsation from \sgr\/ was missing (\citealt{2020ATel13699....1Z, 2021NatAs...5..414K,2021arXiv210604821T, 2020ATel14074....1G, Bailes, Rea}).

Among the $\sim$ 30 magnetars known (\citealt{2014ApJS..212....6O}) \footnote{\url{https://www.physics.mcgill.ca/~pulsar/magnetar/main}}, only five of them have shown radio pulsations (Swift J1818.0-1607, SGR 1745-2900, PSR J1622-4950, XTE J1810-197, 1E 1547.0-5408). A monitoring campaign of \sgr\/ with the Five-hundred-meter Aperture Spherical radio Telescope (FAST) was carried out during the April 2020 outburst. A radio burst was detected on 30 April 2020 21:43:05 (coordinated universal time, UTC), two days after FRB~20200428. We then detected radio pulsations in 464 rotation cycles from \sgr\/ with 563 pulses (Appendix Table \ref{tab:bursttab}), making it the sixth member of the rare radio magnetar population. 

The primary objective of this study is to summarize the properties of the pulsed emission observed during the same epoch in October 2020, based on the radio detections made by FAST and the X-ray detections made by NICER and SWIFT. The second primary goal is to use this comprehensive analysis of combined radio and X-ray detections to address critical pieces of the astrophysical puzzle such as the following.

\begin{itemize}

\item{What is the intrinsic correlation between the behaviour of the magnetar radio and X-ray radiation?}

\item{Is there a different origin for magnetar radio emission and canonical pulsars?}

\item{Can there be a unified origin story to understand these pulses and the first Galactic FRB?}

\end{itemize}

The structure of this work is as follows: Section 2 describes the observations and data analysis procedures; Section 3 presents the pulse catalog and comprehensive radio pulsation analysis from our sample; Section 4 studies the bimodal behaviors of the X-ray hardness ratio, correlated with the activity level of radio emission; Section 5 discusses the radiation mechanism of the radio active period accompanied by the X-ray spectral hardening; Finally, our conclusions are summarized in Section 6.

\section{Observation and data reduction}
The depth and cadence of FAST monitoring allow us to do a detailed study of \sgr\/. Part of the FAST data were also presented in \citealt{zhu2020, zhu2023}, which presented a detailed study of the pulsar timing properties, but with completely different analysis. In this paper we focus on the unique X-ray spectral hardening in \sgr, as it is the only confirmed radio-pulsation-related spectral change of magnetars on a months time scale.

\subsection{FAST observation of \sgr\/ and data reduction.}
\begin{figure*}
\centering
\includegraphics[scale=0.5]{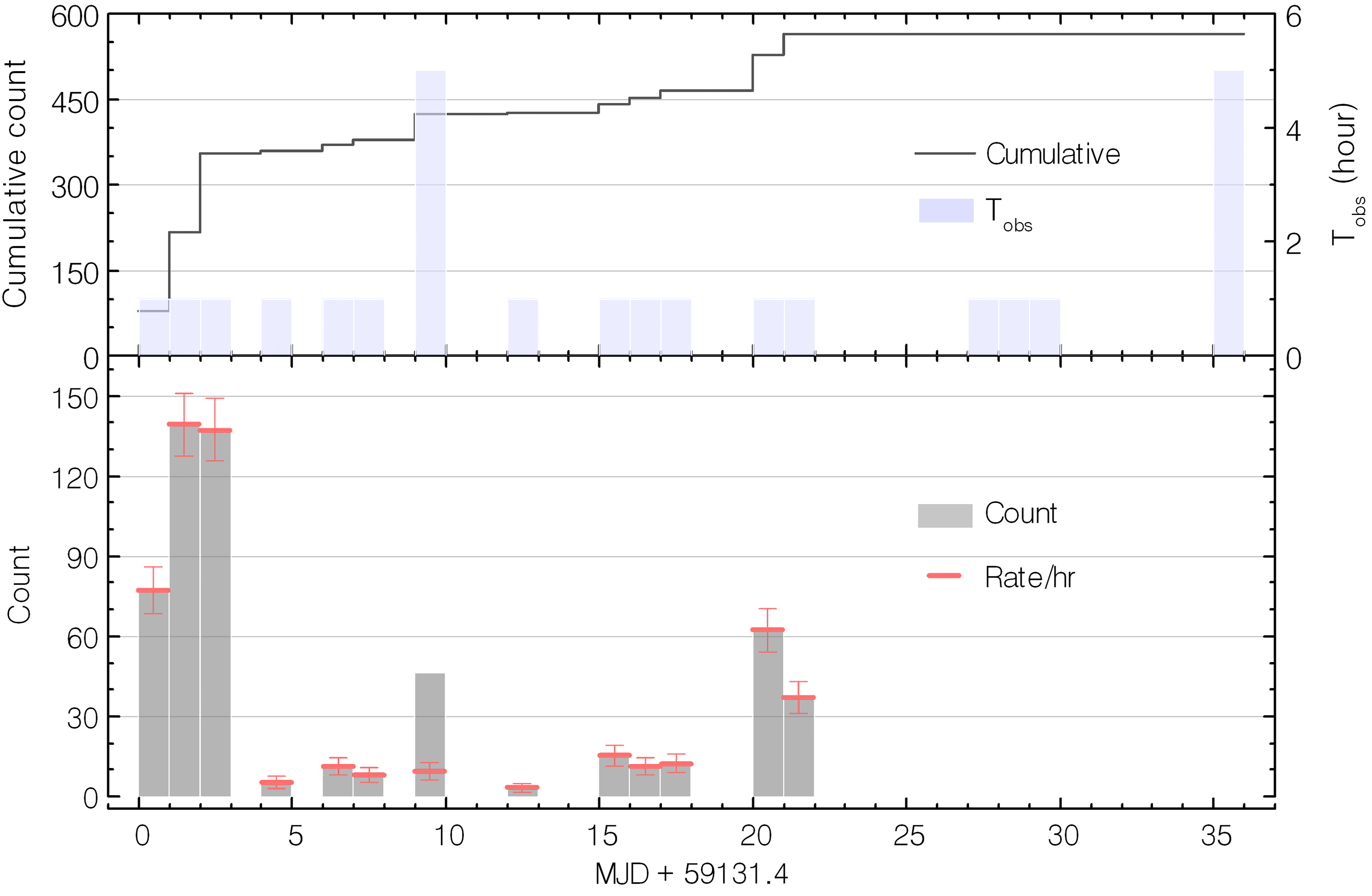}
\caption{Detected radio pulse and temporal burst rate distributions of SGR J1935+2154 during the observation campaign. Upper panel: the duration of each observation session (blue bar, right axis) and the cumulative count distribution of the pulses (solid line, left axis). Bottom panel: the count (grey bar) and rate (red line) of the radio pulses detected.}
\label{rate}
\end{figure*}

The FAST observation of \sgr\/ was conducted for 45 hours mainly in the following two sessions: (i) from 15 April 2020 to 20 May 2020 (UTC); (ii)  {from 9 October 2020 to 7 November 2020 UTC, through the FRB key project of FAST. The distribution of radio pulse and temporal burst rates detected during the observation campaign of SGR J1935+2154 is shown in Figure~\ref{rate}.} The central frequency was 1.25 GHz, spanning from 1.05 GHz to 1.45 GHz, including a 20 MHz band edge on each side. The average system temperature was 25 K. The recorded FAST data stream for pulsar observations is a time series of total power per frequency channel, stored in PSRFITS format  (\citealt{2004PASA...21..302H}) from a ROACH-2 based backend, which produces 8-bit sampled data over 4k frequency channels at 49 $\mu$s cadence. We searched for radio pulsations with either a dispersion signature or instrumental saturation in all FAST data collected during the observational campaign. 

Four types of data processing were performed: I) folding the data according to the ephemerides derived from the X-ray data, II) dedicated single pulse search, III) baseline saturation search, and IV) Artificial Intelligence (AI) aided data filtering.

I. Exploit the X-ray ephemerides to fold for radio pulsed emission in FAST data:
We analyzed the radio data of \sgr\/ using the ephemerides derived from the X-ray data as well as the X-ray light curve during the time span covered by the FAST observation. We separately folded the FAST data from the whole observation as well as those dedicated time spans of detected single pulses.

II. Dedicated single pulse search:
We created de-dispersed time series for each pseudo-pointing over a range of DMs from 200 - 450 pc cm$^{-3}$, which should cover all the uncertainties in a semi-blind search. The step size between subsequent trial DMs ($\Delta$DM) was chosen such that over the entire band t($\Delta$DM) = t$_{channel}$. This ensures that the maximum extra smearing caused by any trial DM deviating from the source DM by $\Delta$DM is less than the intra-channel smearing. We used the above dedicated search scheme to de-disperse the data. Then we used 14 grids of matching filters. The grids were distributed in logarithmic space from 0.1 ms to 30 ms. A zero-DM matched filter was applied to mitigate RFI in the blind search. All the possible candidate plots generated were then visually inspected. Most of the candidates were RFIs, and no pulsed radio emission with dispersive signature was detected with a S/N $>$7.

III. Saturation search:
We understand that FAST would be saturated if the radio flux is as high as hundred-Jansky to mega-Jansky. We therefore also searched for saturation signals in the recorded data-set. We looked for the epoch in which 50$\%$ of channels satisfy one of the following conditions: 1) the channel is saturated (255 value in 8-bit channels), 2) the channel is zero-valued, 3) the RMS of the bandpass is less than 2. We did not detect any saturation lasting $>$0.5 s, hence excluded any saturation associated during the observational campaign.

\begin{figure}
    \centering
    \includegraphics[width=1\textwidth]{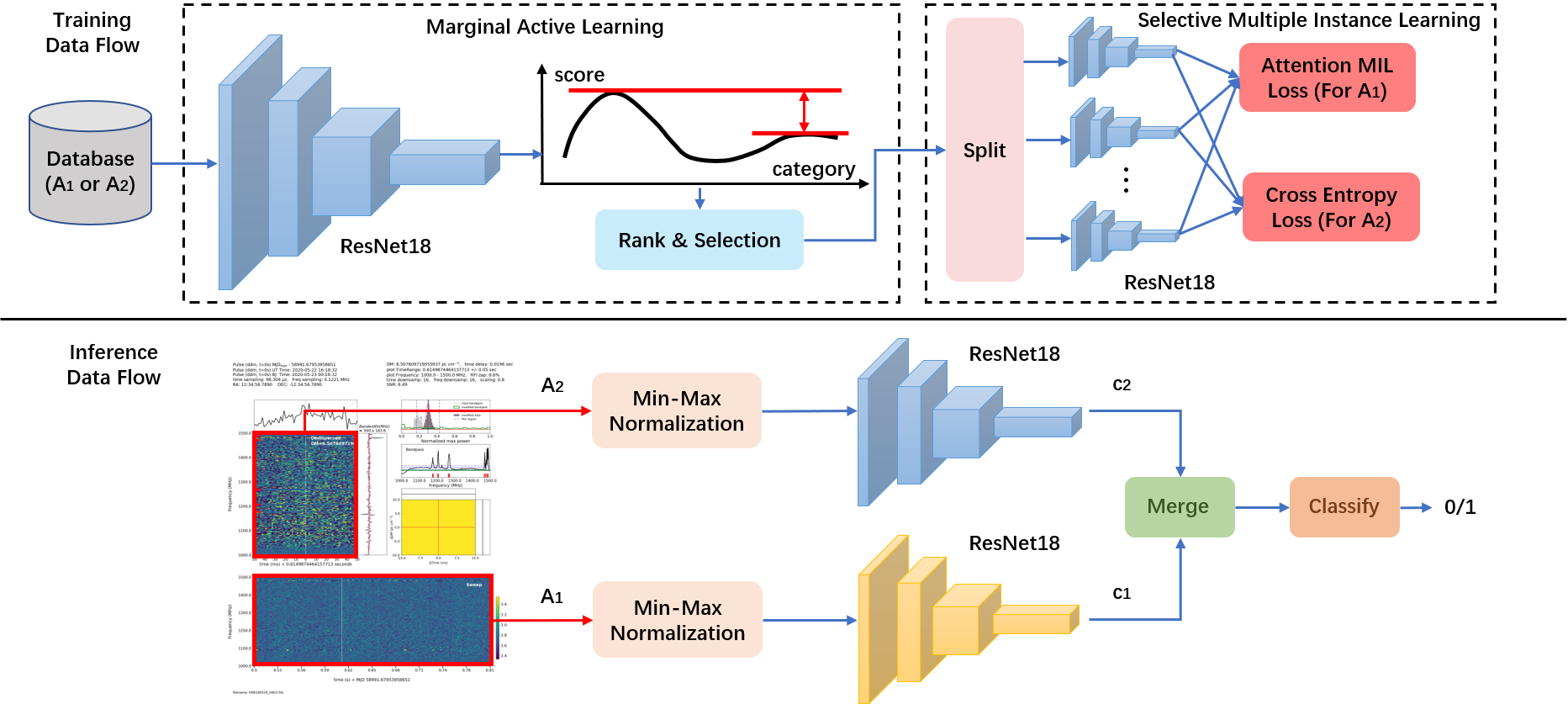}
    \caption{Data flow of AI aided observation data filter. Upper panel: the training data flow to guide the Deep Neural Network (DNN) to fit the joint distribution of positive and negative training samples from FRB~20121102A~ (\citealt{Li2021}). Bottom panel: the inference data flow when taking well trained classifier for online data filtering.}
    \label{classifier}
\end{figure}

IV. AI aided data filtering:
To further reduce the noise and interference in observation, we designed a learning-based filtering framework (Figure~\ref{classifier}) for efficient data selection. In particular, for each observed data with arrival time $t_a$ and trial DM, two types of input are extracted from the observation (1) A time-frequency array $A_1$ cropped from original observation signal in the frequency window $1.05$ GHz to $1.45$ GHz, and time window starting from arrival time $t_a$ and ending at $t_a + \Delta t_1$. The endurance $\Delta t (s)$ can be estimated with the trial DM:
\begin{equation}\label{eq:delay}
    \Delta t_1=4148808.0 \times DM \times \left(\frac{1}{(1450/MHz)^2} - \frac{1}{(1050/MHz)^2}\right)/1000 \, \, \, .
\end{equation}
(2) A time-frequency array $A_2$ cropped from de-dispersed observation with the same frequency window and the time window is centered at $t_a$ and its duration $\Delta t_2 (s)$ is set as the time of $50$ sampling bins.

Both arrays are preprocessed with min-max normalization before sending to AI model. We take ResNet-18~ (\citealt{he2016deep}) as classification model to map the input array into a normalized confidence vector, indicating whether the input array belongs to a potential candidates. For each types of input, we train a classification model respectively. After training process, these two models are deployed to predict the corresponding confidence $c_1, c_2$ for input $A_1$ and $A_2$ respectively. Finally, the confidence score is merged as
\begin{equation}
    c = 0.08 c_1 + 0.92 c_2  \, \, \, .
\end{equation}
If $c>0.1$, the observation data will be sent to visual inspection, otherwise it will be discarded.

To train the classification network, we take the observation from FRB~20121102A~ (\citealt{Li2021}) as training data, which are de-dispersed and visualized for manual labeling. {If the visual observation shows strong visual feature and corresponds to the correct astronomical coordinates with a matching position error HPBW$\sim$3 arcmin  (\citealt{jiang2019}), it is labeled as $1$ (positive) to denote this is a positive training sample, otherwise it is labeled as $0$ (negative). To control the balance between positive and negative training samples, we adopt the active learning strategy of Selective-by-Distinctive-Margin (SDM)~ (\citealt{xie2022learning}) by evaluating the distance from a data sample to different categorical clusters for data selection.} When training classifier with $A_2$, the standard cross entropy is utilized as training loss. To train model with larger array of $A_1$, we take the multiple instance learning framework~ (\citealt{pmlr-v80-ilse18a}) to train the classifier on multiple subarrays of $A_1$. {We then calibrated the noise level of the baseline with noise CAL injection and then measured the intensity values of the detected significant pulse profiles, giving the flux measurement and uncertainties for the detected pulses.}

\subsection{NICER and Swift data analysis}
To study the outburst evolution, we analysed all NICER observations on \sgr\/ recorded along 2020. The first NICER observation (obs ID 3020560101, on April 28) is within the burst storm of the source (\citealt{Younes2020}), thus it was excluded from the study of outburst evolution, which is mainly focused on the persistent emission. A bright dust-scattering halo was detected around \sgr\/ in the Neil Gehrels Swift Observatory (Swift-XRT) observation on April 27. Shortly it decayed to nearly background level on April 28 (\citealt{Mereghetti2020}). Thus, excluding the first NICER observation from our analysis also eliminated the possible contamination from the dust-scattering halo. NICER observational data was processed following the NICER Data Analysis Guide\footnote{\url{https://heasarc.gsfc.nasa.gov/docs/NICER/data_analysis/NICER_analysis_guide.html}}. NICERDAS version 8c incorporated in the HEASOFT version 6.29c was used in the NICER data analysis to process observation data from level 1. The source and background spectra were produced using nibackgen3C50 {(see the NICER Data Analysis Guide in footnote)}{\footnote{\url{https://heasarc.gsfc.nasa.gov/docs/nicer/data_analysis/nicer_analysis_guide.html}}}, {adopting the 2020 background model.} NICER response (RMF) and effective area (ARF) files were produced using NICERrmf and NICERarf.

To increase statistics, several nearby spectra from individual observations were combined to have at least 5000 events. Then the spectra were grouped to have at least 50 counts per energy channel using {\sc grppha}. The hydrogen column density is fixed at $2.4\times10^{22}$ $\rm cm^{-2}$ following previous studies (\citealt{Younes2020}), which is also consistent with  (\citealt{CotiZelati2018}). Considering the large absorption and reduced NICER sensitivity at high energies, we limit our spectral analysis of \sgr\/ in the range of 1-5 keV. All spectra could be well fitted using an absorbed blackbody plus powerlaw model, adopting the Tubingen-Boulder model (tbabs) for interstellar absorption with the abundances adopted in this study (\citealt{wilms2000}). Because of the fitting energy range and statistics, the photon index was fixed at 1.2 (\citealt{Borghese2020,Ersin2020}).
The NICER observations are listed in Table \ref{table:nicer}.

Swift-XRT has monitored the 2020 X-ray outburst of \sgr\/. To avoid the burst storm and focus on the persistent emission of the source, we excluded Swift observations before April 29. During the monitoring campaign, Swift-XRT has been working in both Photon Counting (PC) and Window Timing (WT) mode. For a lower background level, only data in PC mode were used in this paper. 
The Swift observational data was processed following the dedicated pipeline \footnote{\url{https://swift.gsfc.nasa.gov/analysis/}}. We selected PC data with event grades 0–12. A circular region centered on \sgr\/ with a radius of 30 pixels (1pixel=2.36 arcsec) were adopted to extract the spectrum of source.  Background spectrum were extracted from  source-free region with a same size. Spectra were combined for a higher statistics and analyzed in the range of 1-10 keV, adopting a similar method as in NICER data.
The Swift-XRT observations are listed in Table 3.

\section{{Comprehensive radio pulsation analysis}}
We measured the peak flux density, equivalent pulse width and fluence etc. for each pulse (see Appendix Table \ref{tab:bursttab} for the full catalog). The timeline of the pulse fluence during the FAST observation campaign, combined with reported measurements of upper limits at various radio instruments/frequencies (\citealt{Bailes}) are shown in the upper panel of Figure \ref{MJD}. The average fluence is 0.03$\pm$0.01 Jy ms, which is 7-8 orders of magnitude lower than that of FRB~20200428. {The pulse rate peaked at 139 hr$^{-1}$ on 10 October (MJD 59132) and at 137 hr$^{-1}$ on 11 October (MJD 59133), which means that the fraction of rotation periods with pulse detection is $\sim$10\% of the total number of rotation periods during the corresponding observation time (see Figure \ref{rate}).} The pulse rate then precipitously dropped by an order of magnitude.

\begin{figure*}
\centering
\includegraphics[scale=0.6]{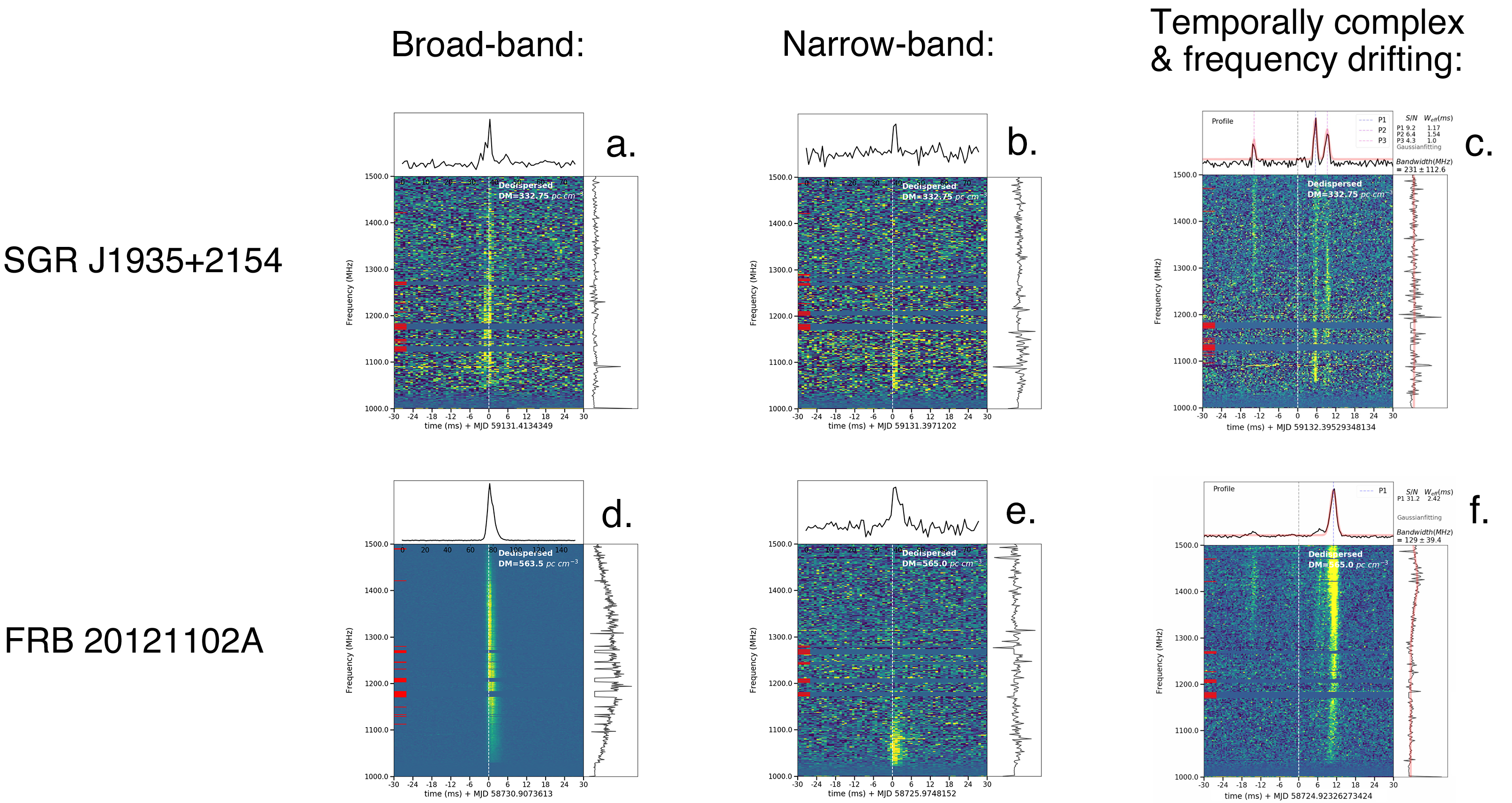}
\caption{{Three archetypes of pulse morphology in the time-frequency waterfall plots. Rows from left to right: Broad-, narrow-band, and temporary complex/frequency drifting dedispersed dynamic spectra of the radio pulse morphology of SGR 1935+2154 as detected by FAST on Oct. 2020, compared to repeating FRB~20121102A (also with FAST, between 29 Aug. and 4 Sep. 2019).
The archetype spectra of \sgr\/ show an interesting morphological similarity with repeating FRB. The band-averaged time series and time-averaged spectra are shown on the top and right sides of panel (c) and (f), respectively, with the Gaussian-fitted models overplotted by red-solid lines.}}
\label{dyn}
\end{figure*}

Radio pulses from \sgr\/ have shown unique time-frequency properties, among which broad \& narrow-band emission and {an interesting sub-pulse time–frequency upward- or downward-drifting morphology}. We identify three observed archetypes of pulse morphology in \sgr\/ (Figure \ref{dyn}, upper panel a, b, c). Narrow-band and frequency drifting emissions {were} seen in FRB morphology  (\citealt{Pleunis2021}) (e.g. FRB~20121102A, Figure \ref{dyn}, panel d, e, f), but rarely observed in normal pulsars, and never detected in magnetars before \sgr\/. The drifting rate of the so-called downward frequency drifting morphology in the dynamic spectrum was defined by Hessels et al.  (\citealt{Hessels2019}).

\subsection*{{i)} DM variation}
To quantify the specific morphological characteristics of the pulsation, the optimum dispersion measure (DM) should be determined. We utilized the incoherent dispersion technique using the single-pulse-search tool in PRESTO~ (\citealt{Ransom2001}). De-dispersed pulse profiles were created for each DM trial between 300 to 350~pc~cm$^{-3}$ with a step size of 0.05~pc~cm$^{-3}$ to maximize the ‘local contrast’ of the dedispersed pulse profile. Gaussians (multiple when necessary) were fitted to the structure function of the profiles $I(DM,\ t)$. The derivative of each Gaussian was then squared. For multiple components, the square of the derivative of the pulse profiles was summed. The optimized DM$_{obs}$ was then determined though the maximization of the area under the squared derivative profiles, thus maximizing the structures in the frequency integrated burst profile:
\begin{equation}
     DM_{obs} = \underset{DM}{\arg\max}\ \left\{\sum_{t} \left( \frac{dI(DM,\ t)}{dt} \right)^{2} \right\}
\end{equation}
the FWHM of the fitted DM is taken as the uncertainty. The optimal DM of the pulses is finally constrained to be 332.7$\pm$0.1 pc cm$^{-3}$ between MJD 59131 and MJD 59167.

{The DM distribution of these pulses exhibits no significant time-evolutionary feature. In order to test the reliability of the no-evolutionary trend of DM, we divided the DM measurements into two time-epochs according to the dates of the event and generated mock DM values in each bin, based on the mean and the standard deviation of the measured DMs in the respective time bins. The null hypothesis was then tested based on the generated DMs under the assumption that DM does not change over time. For each set of generated DMs, a slope was fitted. Based on 20,000 trials, the central value of the slope distributions is zero with a $\sigma$ of $\sim$0.004 pc cm$^{-3}$ yr$^{-1}$ for both time-epochs. The rapid fluctuations in DM values between pulses can be attributed either to the low flux density that does not allow for a good fit to the pulse profiles, or to external causes such as turbulent motion of clumps or filaments in the surrounding pulsar wind environment.}

\subsection*{{ii)} Energy distribution}
\begin{figure*}
\centering
\includegraphics[scale=0.45]{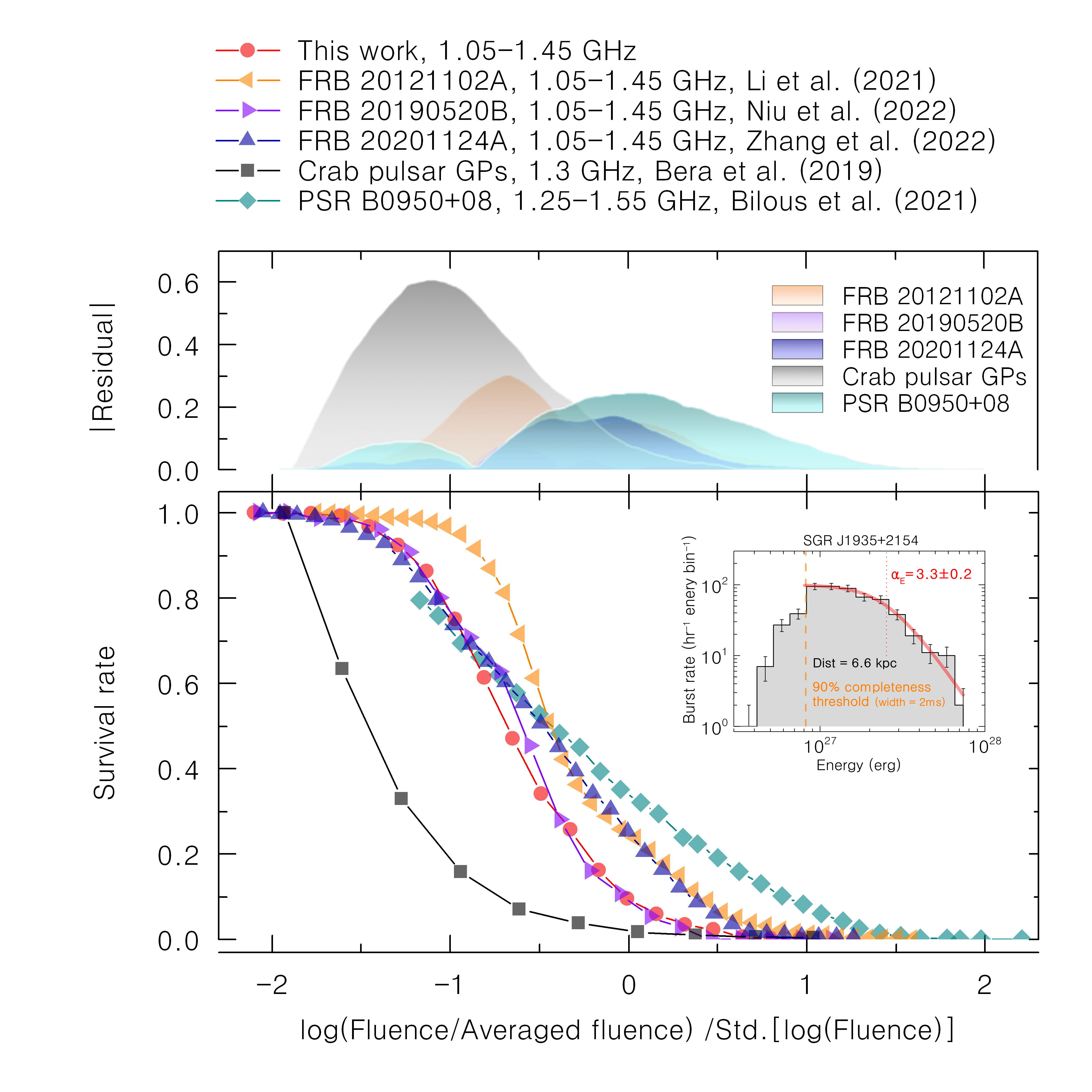}
\caption{Comparison of the single-pulse fluence cumulative distribution for classical pulsars, pulsar giant pulses (GPs), FRBs and \sgr\/. The collection of fluences are normalized both by their averaged fluences and standard deviation, respectively. The subplot is the pulse rate distribution of the isotropic equivalent energy at 1.25 GHz for \sgr\/ pulses. A single power-law fit for pulses above a certain threshold E $\geq$ 2.5$\times$10$^{27}$ erg is shown in red solid line, and the 90$\%$ detection completeness threshold is shown with the yellow dashed line.}
\label{survivalRate}
\end{figure*}

The pulse rate versus pulse energy distribution of \sgr\/ shows a broad bump (Figure~ \ref{survivalRate}). We test the hypothesis that the energy distribution can be described by a power-law function $N(E) = N_1 E^{-\alpha_{E}}$ in a certain energy range $E_i\ \leq E \leq\ E_f$:
\begin{equation}
N_1 = \frac{N_{ev}(1-\alpha_{E})}{E_f^{1-\alpha_E}-E_{i}^{1-\alpha_E}},
\end{equation}
where $N_1$ is the normalization constant, and $N_{ev}$ is the total number of pulses included. This definition is only valid for the index $\alpha_{E}$ $\neq$ 1, while the integral takes the form of $\propto$ ln(E) for $\alpha_{E}$ = 1. The energy function is consistent with a power law with $\alpha_E = 3.3\pm0.2$ and a central energy around $4.8 \times 10^{27}$~erg for the energy range $E \geq$ 2.5$\times$10$^{27}$ erg with a coefficient of determination $R^2 = 0.928$.

\begin{figure*}
\centering
\includegraphics[scale=0.9]{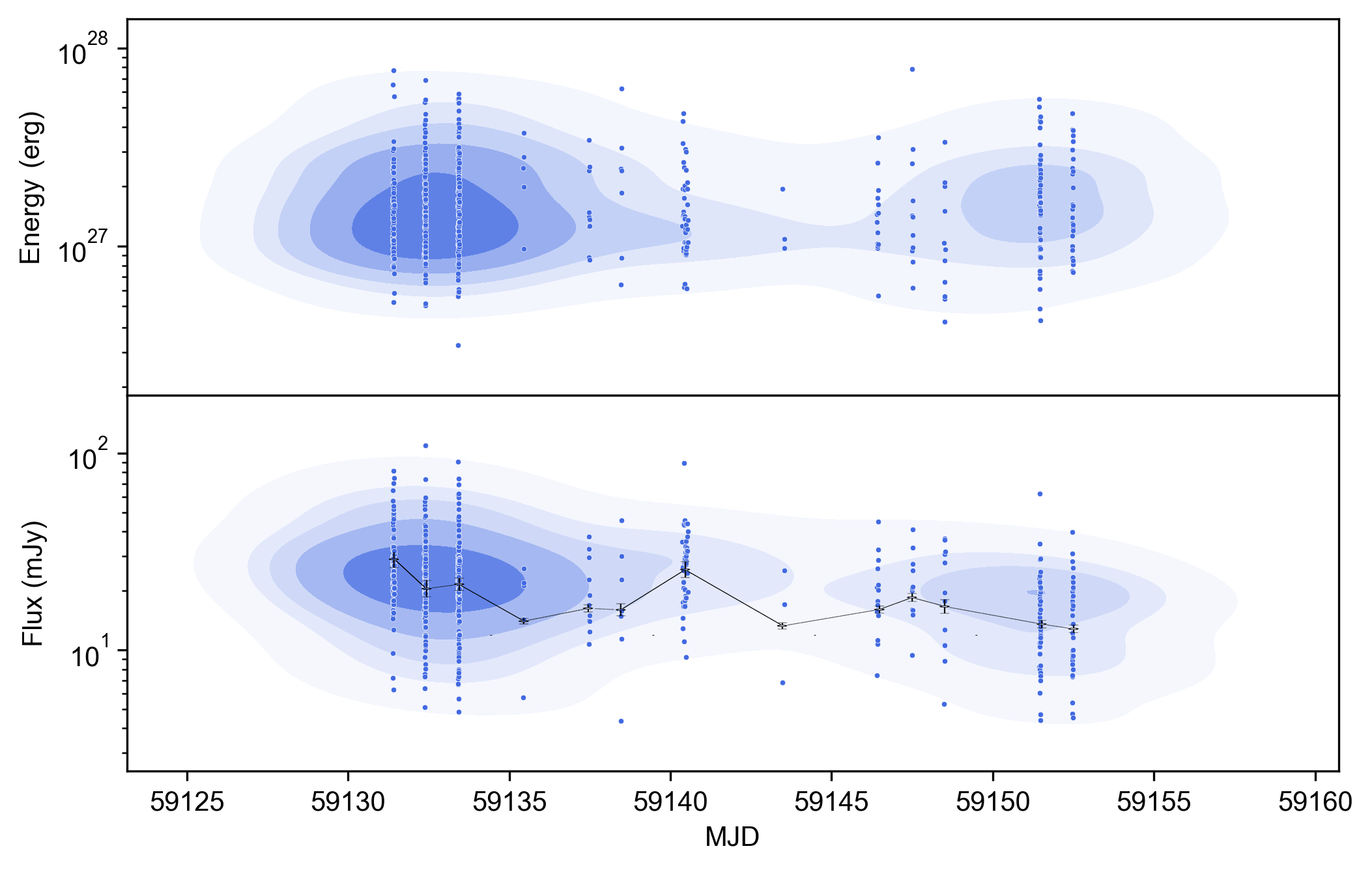}
\caption{The 2D kernel density estimation (KDE) of the pulses detected from \sgr\/. Upper panel: the isotropic energy KDE (blue contours) assuming a distance of 6.6 kpc. Lower panel: the flux KDE (blue contours) and the average value for each observation session (black points). }
\label{kde}
\end{figure*}

{This distribution is time dependent and shows a bimodality along time.} There are days with significantly higher pulse rate averages, although the averaged pulse energy variations is small. {The high energy pulses are more concentrated before MJD 59140 (Figure~\ref{kde}). There is no significant correlation between the averaged pulse energy with its standard deviation for each day and the pulsed event rate.} The fewer events after that epoch may signify that the bimodality is due to time-dependent lensing, but more plausibly it might be an analog of ``mode changes'' commonly seen in long period radio pulsars where pulse components change their relative amplitudes and occurrence rates. Further observations can distinguish between these possibilities.

We measured the peak flux density, pulse width and fluence, then calculated the isotropic equivalent energy of each pulse based on the latest distance estimation to \sgr\ (6.6$\pm$0.7 kpc (\citealt{zhou2020})) at 1.25 GHz. The derived pulse energies span more than one order of magnitude, from 5$\times$10$^{26}$ erg to $\sim$10$^{28}$ erg. The normalized cumulative distribution of the pulse rate for \sgr\/ (hereafter defined as ``survival rate'' in Figure \ref{survivalRate}) can be well fitted by a single threshold power-law function with index of 3.3$\pm$0.2 after considering the instrumental detection completeness bias. The index of \sgr\/ is similar to known pulsars or pulsar giant pulses (GPs) with a steep slope, and is different from that of FRBs, which on average is more shallow. We compared the pulse fluence cumulative distributions of classical pulsars (\citealt{Bilous2022}), GPs (\citealt{Bera2019}), repeating FRBs (FRB~20121102A (\citealt{Li2021}), FRB~20190520B (\citealt{Niu2022}) and FRB~20201124A (\citealt{Zhang2022})) and \sgr\/. The fluences are normalized both by the average and standard deviation of fluence in logarithmic space (Figure \ref{survivalRate}). The distribution of \sgr\/ is clearly separate from the Crab pulsar GPs and PSR B0950+08, but {tend to} follow the trend of FRB~20190520B and is {likely} consistent with FRB~20201124A in the lower end of fluence.

\subsection*{{iii)} Morphology}
The pulsed radio emission from magnetars is usually characterized by a flat radio spectral index ($S_\nu \propto  \nu^{0.5})$ and large variabilities both in flux density and pulse profile (\citealt{Camilo2006, Kramer2007}).

\begin{figure*}
\centering
\includegraphics[scale=0.5]{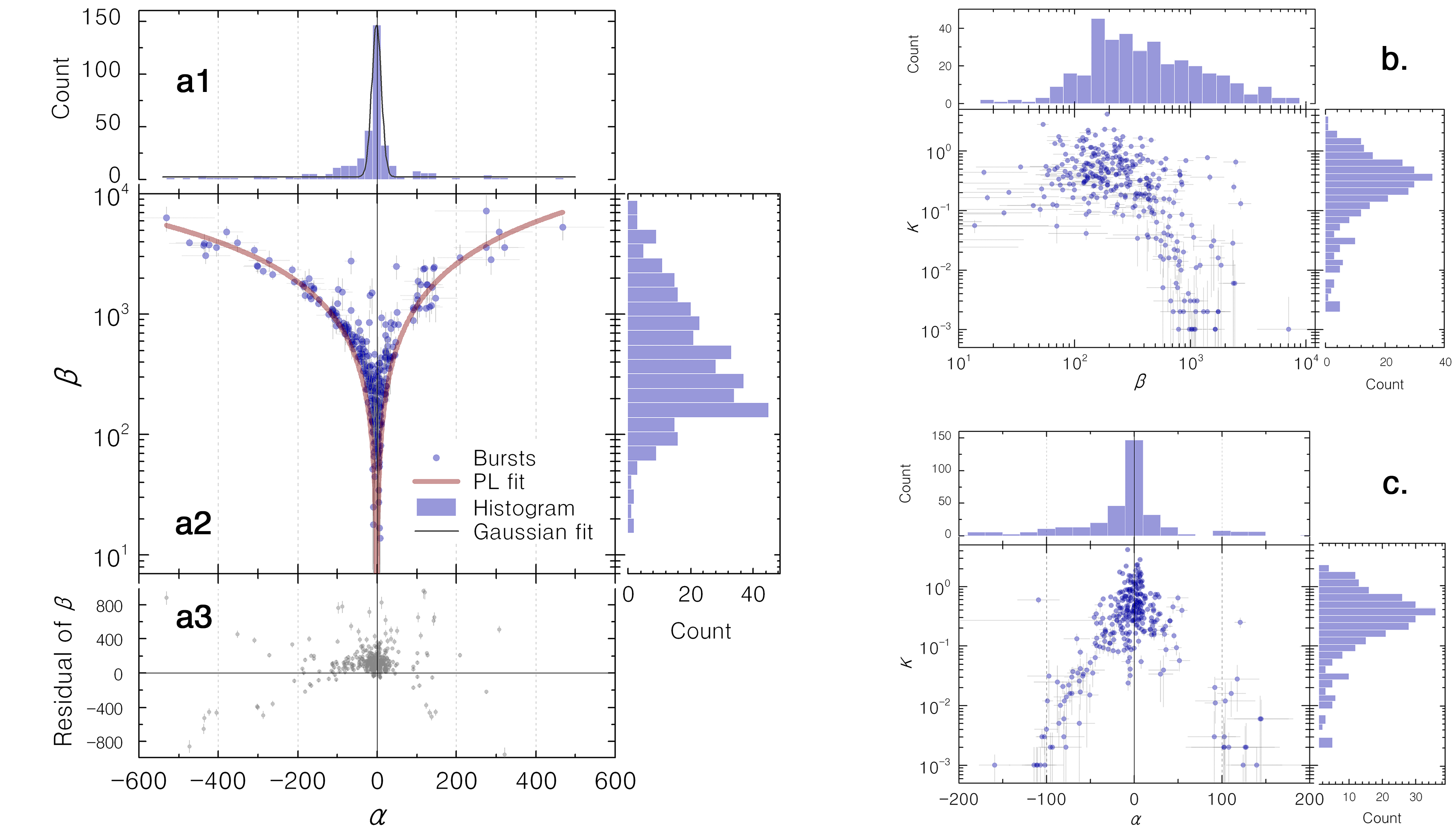}
\caption{The distributions of the morphological fitting parameters $\alpha$, $\beta$ and $\kappa$. In each of the lower left plots of each panel, the points represent pulses with S/N $>$10 from the set of 563 pulses detected. {The error-bars are calculated at the 68$\%$ confidence level. In each panel, the top plot is the histogram of the x-axis data, and the right plot is the histogram of the y-axis data. Panel (a): $\alpha$ vs $\beta$, a clear separation of the narrow- and broad-band populations suggests $\beta$ is a discriminating parameter between them. Two-component PL distribution is separately fitted in red solid line (panel a2), and the fitting residual is in the bottom panel a3). Panel (b) shows $\kappa$ vs $\beta$, and Panel (c) indicates $\kappa$ vs $\alpha$.}}
\label{morphology}
\end{figure*}

{We described the frequency band-pass of pulses with signal to noise ratio (S/N) $\geq$10 using the phenomenologically power-parabola function  (\citealt{Thulasiram2021})
$I = \kappa(\frac{\nu}{\nu_0})^{\alpha+\beta \log(\frac{\nu}{\nu_0})}$, 
where $\nu$ is a frequency value between 1.0 and 1.5 GHz of FAST L-band receiver, $I$ is the amplitude (arbitrary units) at a given frequency, $\alpha$, $\beta$, and $\kappa$ are three of the free parameters and $\nu_0$ = 1.25 GHz is the reference frequency.}
A higher $\beta$ indicates narrower frequency band emission. The fitting function would degenerate to a power-law function with $\beta$$\sim$$0$. We estimated the fitting uncertainties through the standard error propagation to avoid over fitting. The distributions of the morphological fitting parameters $\alpha$ and $\beta$ are shown in the panel (a) of Figure \ref{morphology}. The distribution of $\beta$ against $\alpha$ clearly separates the narrow-band and broad-band pulses. Narrow-band pulses (higher $\beta$) tend to have a steeper spectrum (higher $\mid$$\alpha$$\mid$) than broad-band ones. {A power-law (PL) distribution ($\beta$=A$\mid$$\alpha$$\mid^{\gamma}$) is fitted to the two branches ($\alpha$$\leq$0 and $\alpha$$\geq$0) (panel a2) with respective index of $\gamma_l$ and $\gamma_r$. The best-fit $\gamma_l$ and $\gamma_r$ are the same (1.1$\pm$0.02) and the residuals have no obvious structure (Figure \ref{morphology}, panel a3). } Furthermore, the distribution of $\kappa$ against $\beta$ and $\alpha$ (\ref{morphology}, panel b and c) suggests that $\kappa$ is also a discriminating parameter for narrow-band pulses. Narrow-band pulses (higher $\beta$ and $\mid$$\alpha$$\mid$) are less bright (lower $\kappa$) than broad-band ones. The pulses show a similarly observed archetypes of FRB morphology  (\citealt{Pleunis2021}), and have an apparent difference from representative young pulsars.

The radio pulsations' frequency drifting rate of \sgr\/ is $\dot\nu\sim10$ GHz per second. With the observation frequency $\nu_{\rm obs}=1.45$ GHz, it would suggest a radio emission location, following equation 1 in \citealt{yang2020}:
\begin{equation}
R_{\rm radio}\sim\frac{c\nu_{\rm obs}}{\dot\nu_{\rm drift}}=4.35\times10^{9}~\mbox{cm},
\label{driftrate}
\end{equation}
which is within the light cylinder radius $R_{\rm LC}=c/2\pi\nu=15.5\times10^{9}$ cm but far away from the magnetar surface ($R_{\rm magnetar}\sim10^{6}~\mbox{cm}$). This suggests that the radio emission of \sgr\/ is most likely originated from the closed field lines region in the outer magnetosphere through the formation of a current loop originated from the twisting magnetospheric field line (hereafter referred to as "j-bundle"), rather than the polar cap region as in canonical radio pulsars. 

\subsection*{{iv)} Time-domain analysis}

\begin{figure*}
\centering
\includegraphics[scale=0.5]{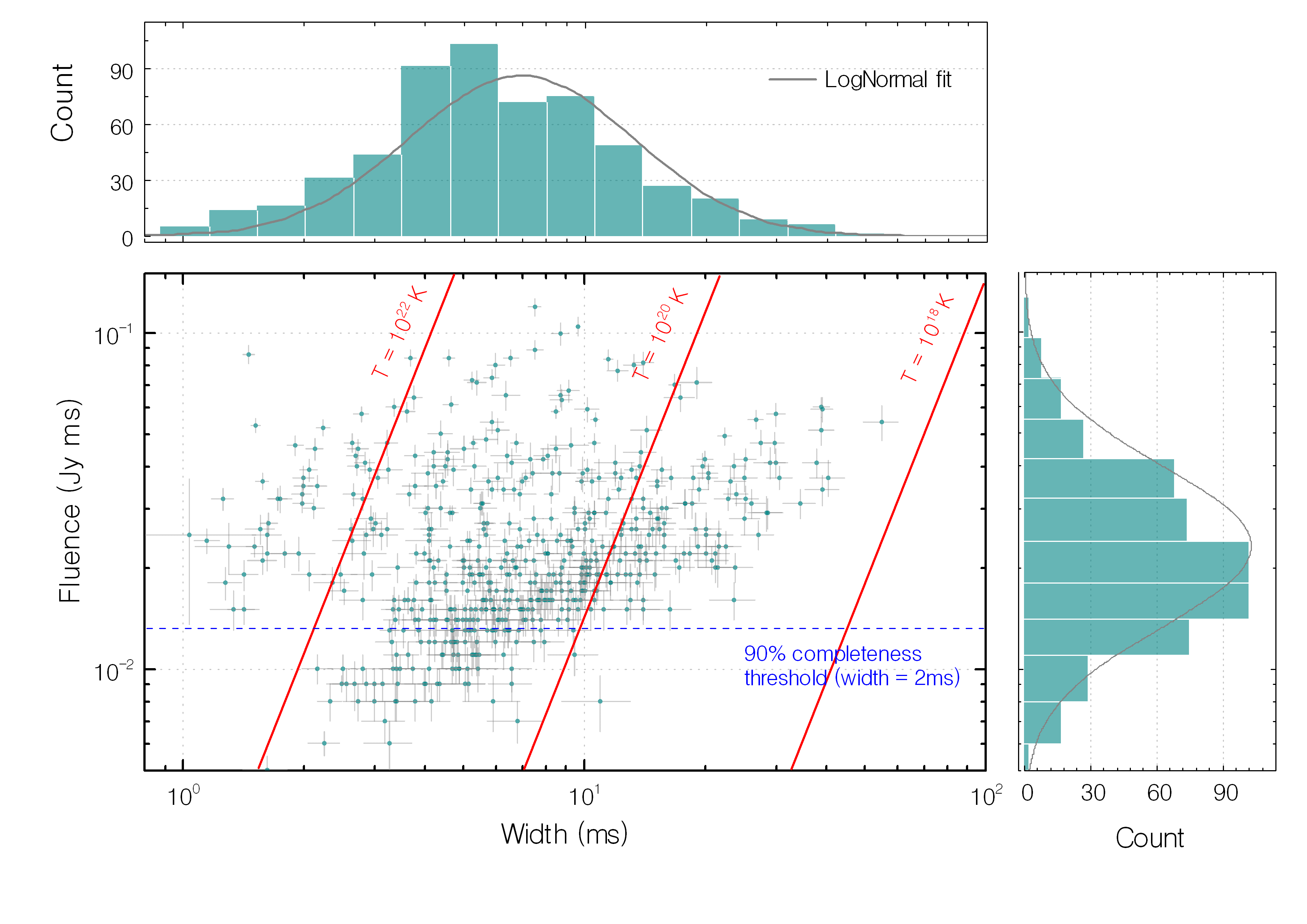}
\caption{The fluence and equivalent pulse width distributions at 1.25 GHz for \sgr\/. The dots indicate all of the 563 detected pulses, the red solid lines represent the equivalent brightness temperatures, and the horizontal blue dashed line is the simulated 90$\%$ detection completeness threshold. In the upper panel, an overall log-normal (LN) distribution is fitted in grey line.}
\label{fluence}
\end{figure*}

{To quantify the specific time-domain characteristics of pulses, we take the pulse profile above the detection threshold as a separate pulse. To be conservative, we take the detection threshold using the 3-sigma noise level of the baseline due to the complicated effect on the actual sensitivity, e.g. the bandwidth limited structure of the pulses and radio frequency interference (RFI) events. On the other hand, we tried to differentiate the pulses by referring to the morphological clustering characteristics of the pulses in the dynamic spectrum.} Figure~\ref{fluence} shows pulse width $W_{\rm eq}$ against fluence distribution. The equivalent width $W_{\rm eq}$ is defined as the width of a rectangular pulse that has the same area as the detected pulse, and the height of the peak flux density is denoted as $S_{\rm peak}$. In our sample, several pulses might be described with multiple components in a single pulsation, if there is ``bridge'' emission (higher than 5 $\sigma$) between pulses for the pulses with a complex time-frequency structure. This results in some pulses having overestimated equivalent widths. The computed $W_{\rm eq}$ range from 0.9~ms to $\sim$40~ms, consistent with a log-normal distribution centered around $\sim$6~ms. This is consistent with the known statistical properties of repeating FRBs, and RRAT population (\citealt{Li2021, Maciesiak2011}).

\begin{figure}
    \centering
    \includegraphics[width=0.8\textwidth]{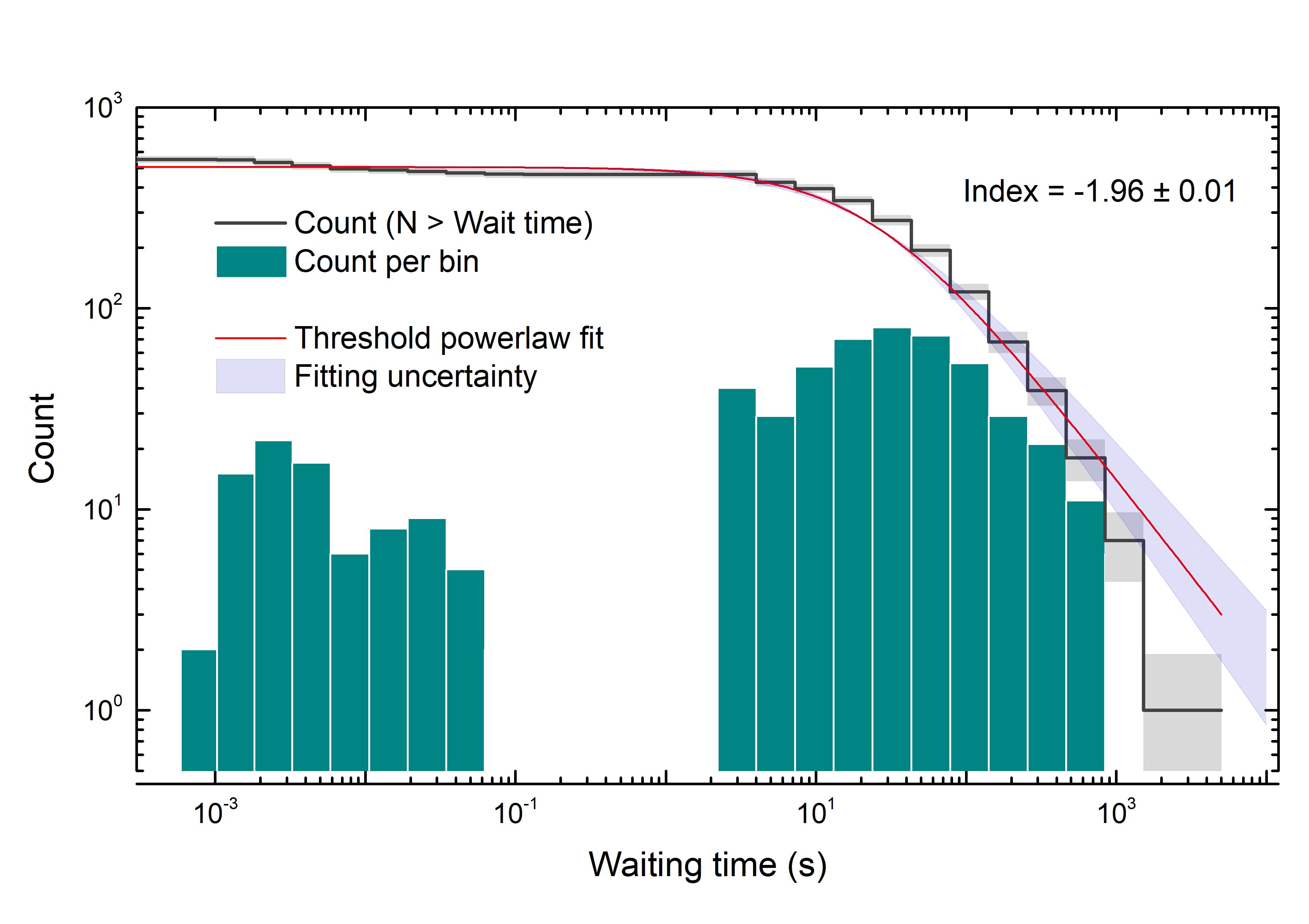}
    \caption{{Waiting-time frequency distribution. We give the differential (green region) and cumulative (solid black line) distribution of waiting-time, respectively. Solid red line represents an overall fit for the cumulative distribution of the pulse rate using a single threshold power-law function with index of 1.96$\pm$0.01, while the shaded region indicates fitting uncertainty.}}
    \label{wt}
\end{figure}

We also studied the statistical property of the waiting time distribution. {The waiting time between two adjacent detected pulsation is $\delta$t = $t_{\rm i+1}-t_{\rm i}$, where $t_{\rm i+1}$ and $t_{\rm i}$ are the corresponding bary-centred arrival times for the ($i+1$)th and ($i$)th bursts. All waiting times were calculated for pulses within the same session to avoid long gaps of $\sim$24~hr. Figure~\ref{wt} shows the occurrence rates (green region) as a function of waiting times as well as its cumulative distribution (solid black line) of the waiting time. The cumulative distribution of the pulse rate can be well fitted from Monte Carlo Markov Chain (MCMC) method using a single threshold power-law function with index of 1.96$\pm$0.01 (solid red line in Figure~\ref{wt}).} {We found similar occurrence frequency distributions of waiting time for FRBs (\citealt{Gourdji2019, FYWang2021, Aggarwal2021}), which also supports the temporal tight correlation between the \sgr\/ pulsations and repeating FRBs.}

\begin{figure*}
\centering
\includegraphics[scale=1]{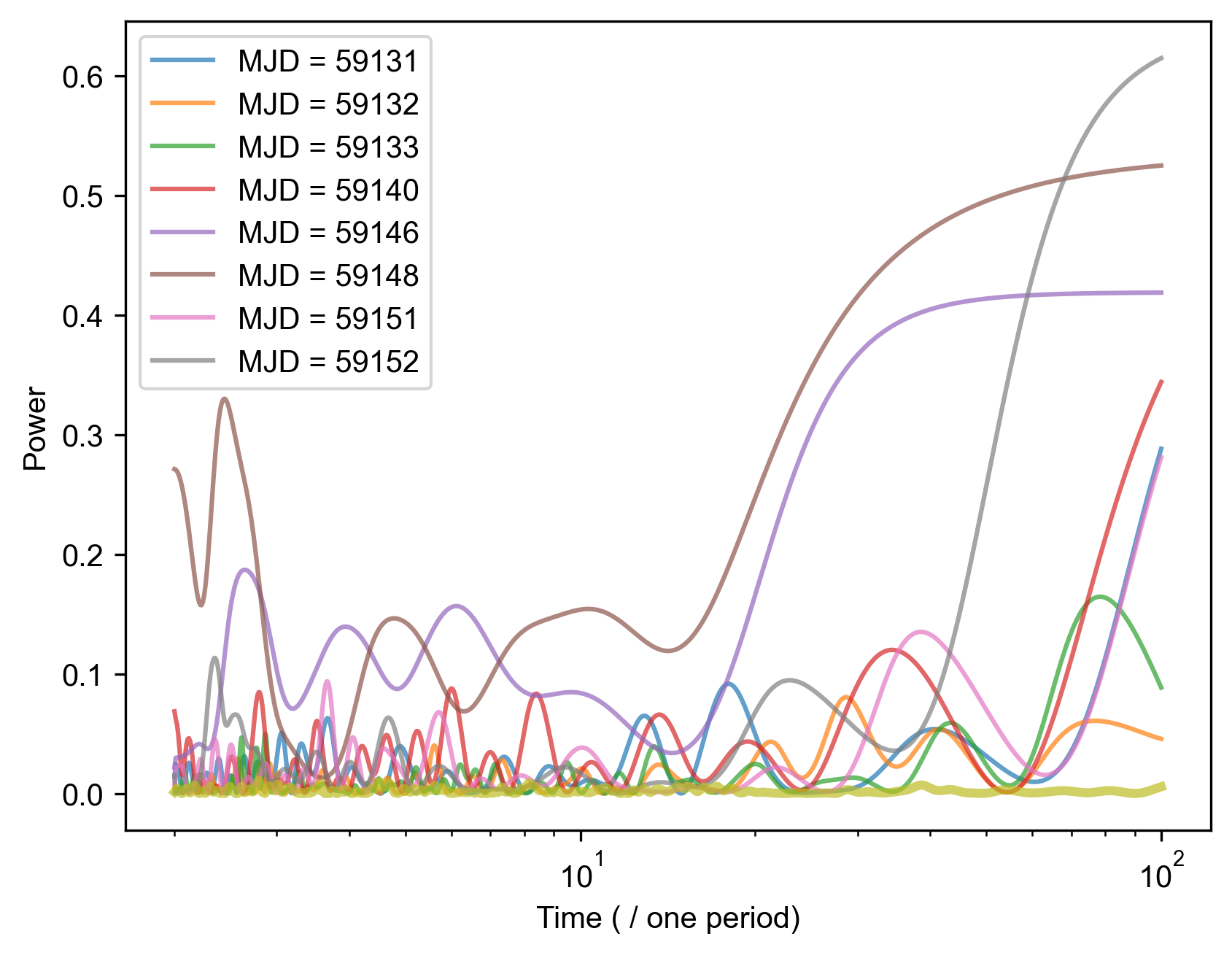}
\caption{Lomb-Scargle periodograms of nulling periods for \sgr\/. Nulling period has been searched from 2 to 100 rotational periods, and no distinct concentration has been found in all of the pulse active phases.}
\label{null}
\end{figure*}

The Lomb–Scargle periodogram method is widely used to identify periodicities in data that are not uniformly sampled. We here apply this method to the nulling of rotation periods to determine whether there is a possible period. If the active phase of nulling does have a period, folding the time intervals of nulling according to amplitude-modulated periodicities would show clustering in burst phase, suggesting a modulation pattern compatible with a rotating beam-carousel model. Periodograms of nulling from \sgr\/ for periods ranging from 2 to 100 are shown in Figure~\ref{null}. No periodicity of the nulling time was found in the power spectrum. Folding with that period does not show any concentration in pulse phase. In addition, we design an experiment in order to test whether the periodicity of Nulling has a possible pulse energy dependence. We divide all the detected pulses into two groups: the full pulse and the bright pulse with S/N $\geq$ 15 are subjected to the periodic search separately, and the power spectra of the two groups are compared. The test results show that neither group reflects any sign of periodicity in nulling, and that lack of periodicity for the pulsation nulling intervals is independent of the grouping of pulse energy sizes. All observing sessions fall in the pulsation active phase of \sgr\/, thus the pulsation nulling intervals also did not reveal any underlying modulation periodicity.

\section{Bimodal behaviors of the X-ray hardness ratio, correlated with the activity level of radio emission}
We carried out a detailed comprehensive pulse analysis on energy, duration and waiting time for \sgr, which shows an apparent difference from representative young pulsars. Previously detected five other magnetars (Swift J1818.0-1607, SGR 1745-2900, PSR J1622-4950, XTE J1810-197, 1E 1547.0-5408), the radio pulsations of which are usually detected shortly after their respective X-ray outbursts. (\citealt{camilo2007a, halpern2008}). At some point, the radio pulsations may disappear without related activities in other wavelengths (e.g. in X-rays), to rise again in other outbursts. The delays between radio pulsations and the onset of X-ray outbursts range from as short as $\sim$35 hours (Swift J1818.0-1607 (\citealt{Karuppusamy2020})) to $\sim$18 days (XTE J1810-197 (\citealt{Caleb2022,Lyne2018})). For SGR 1745-2900 and 1E 1547.0-5408, the delay times are $\sim$5 days (\citealt{Rea2013}) and $\sim$3 days \& $\sim$16 days (\citealt{Bernardini2011,Camilo2009,Lower2023}), respectively. For PSR J1622-4950, the radio bursts were first detected in 2009, and again in 2017, years after the prior closest X-ray observation, during which, however, no X-ray burst was detected, thus rendered the delay time uncertain (\citealt{Levin2010,Anderson2012,Camilo2018}). Out of all five magnetars aforementioned, the X-ray spectral hardening associated with radio pulsation was only observed in SGR 1745-2900 (\citealt{Kaspi2014}; 5 days delay). However, Kaspi et al. argued that such behavior could originate from a different source within the NuSTAR PSF, due to its proximity to the Galactic center. Thus, the X-ray spectral hardening in SGR J1935+2154 is unique: it is the only confirmed radio-pulsation-related spectral change, and has a time scale of months.

\begin{figure*}
\centering
\includegraphics[scale=0.4]{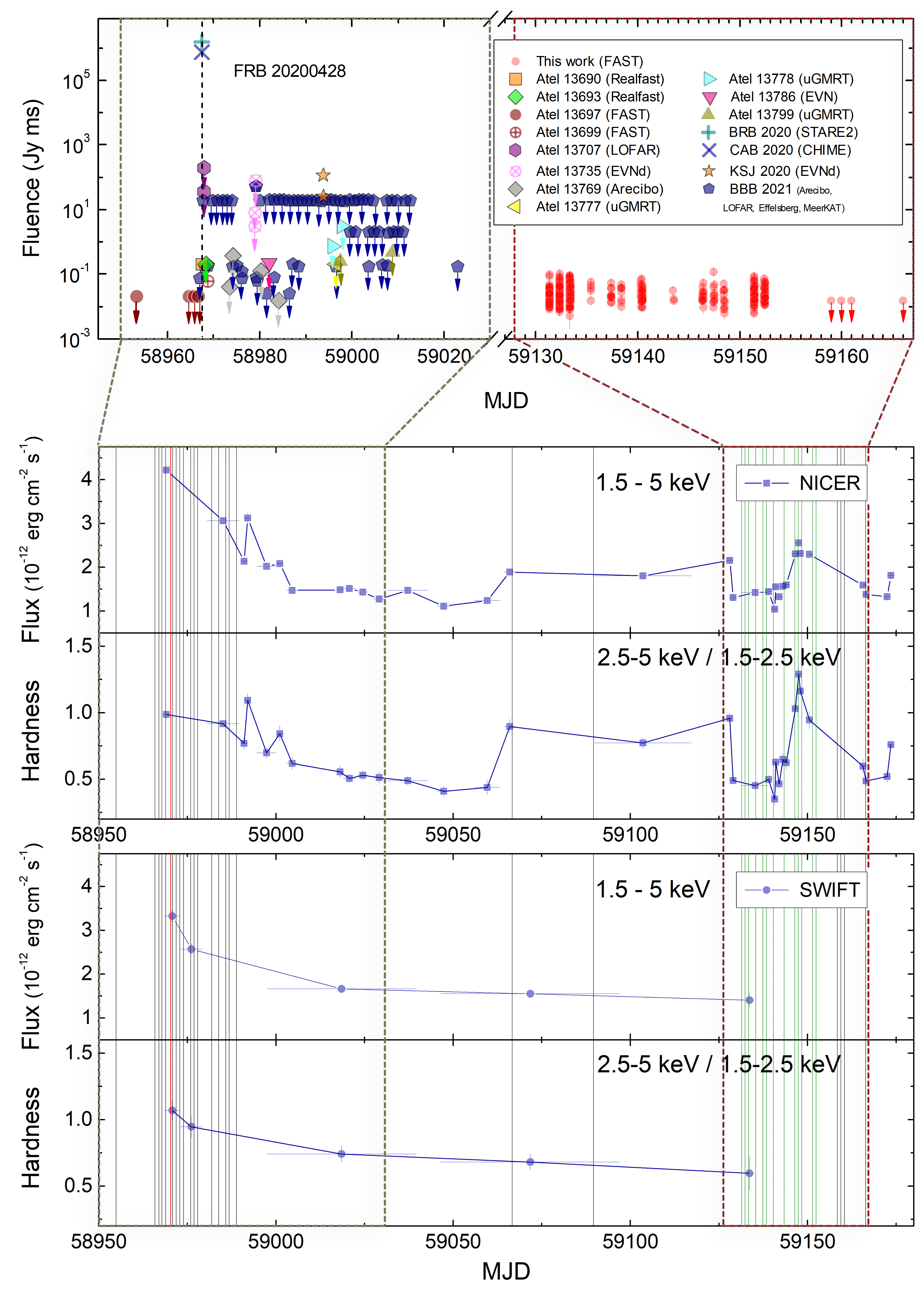}
\caption{Timeline of detected radio pulses and X-ray light curves of SGR J1935+2154 during the observation campaign. (Upper panel) {Temporal pulse fluences or upper limits at various radio instruments/frequencies. The red dots represent all 563 pulses detected by FAST in 464 rotation periods from MJD 59131 to 59167, the downward arrows indicate each upper-limit measurement.} (Bottom panel) From top to bottom: \sgr\/ NICER/SWIFT-XRT light curves in 1.5-5 keV and hardness ratio (2.5-5\,keV/1.5-2.5\,keV) . The solid lines indicate FAST observations with radio flare (red), no radio pulsation (black) and with radio pulsation (green).}
\label{MJD}
\end{figure*}

\sgr\/ has however manifested a different behavior. Radio observations on \sgr\/ are shown in Figure \ref{MJD}, together with the X-ray outburst evolution observed by NICER and Swift-XRT. Its regular radio pulsation was not activated in line with the X-ray outburst, but delayed for several months. The radio active episode of \sgr\/ (green solid lines in Figure \ref{MJD}, lower panels) lasts $\sim$50 days at most, by far the shortest among radio magnetars. During the radio active episode, a significant spectral hardening was detected while the X-ray activity level decayed. The X-ray flux hardness ratio (2.5-5\,keV/1.5-2.5\,keV) observed by combining NICER and SWIFT (Figure \ref{MJD}) underwent a sudden drop around MJD 59140, and then experienced an apparent increase until MJD 59150, reaching the hardest spectrum during the X-ray outburst decay. Dramatic X-ray spectral changes during magnetar outbursts are common, generally starting from a rapid initial hardening and then entering a slow softening as the flux decays to quiescence\footnote{{SGR 1745-2900 has shown spectral hardening during X-ray decay, but could be due to a different source appearing within the NuSTAR PSF considering its proximity to the Galactic center (\citealt{Kaspi2014}).}}  (\citealt{Kaspi2017, CotiZelati2018}).

\begin{figure*}
\centering
\includegraphics[scale=1,angle=90]{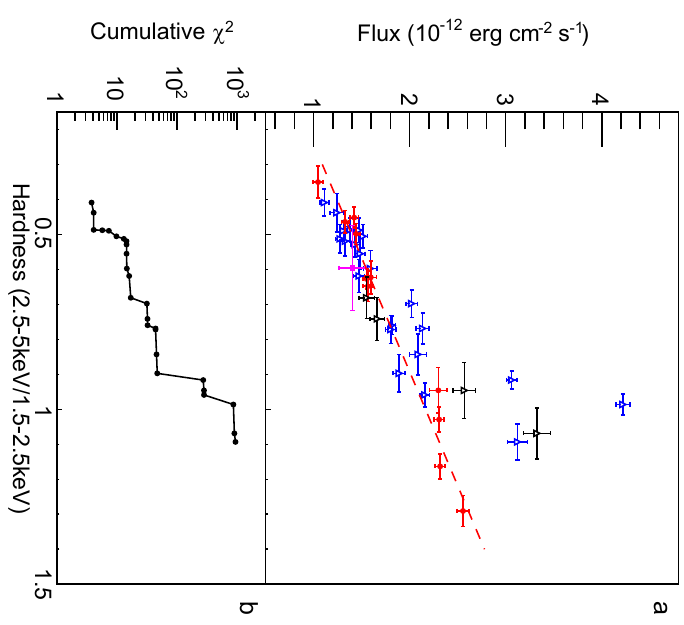}
\caption{
Panel a: Hardness-Intensity Diagram of \sgr\/. The flux is in the 1.5-5 keV band. The red and pink data points are derived from NICER and Swift-XRT observations when SGR 1935+2154 shows radio pulsations, while the blue and black data points represent the rest of the NICER and Swift-XRT observations. The dashed red line represents the linear fitting of the observations when SGR 1935+2154 shows radio pulsations (red and pink data) points. Panel b: cumulative $\chi^{2}$ for observations when SGR 1935+2154 shows no radio pulsations against the linear fitting, showing a trend of deviation.
}
\label{NICERlc}
\end{figure*}

In \sgr\/, we discovered the first confirmed significant spectral hardening while the magnetar outburst decays and with associated radio activity. The Hardness-Intensity Diagram (HID) is a model independent way for tracking the spectral evolution. The panel a of Figure \ref{NICERlc} shows the HID for the \sgr\/'s outburst in 2020. The X-ray data during the radio active period (red (NICER) \& pink (Swift) points) take a different branch from the other outburst decaying period (blue (NICER) \& black (Swift) points). Red and pink data points could be well fitted with a linear function $y=ax+b$, yielding fitting parameters a=1.54$\pm$0.09, b=0.63$\pm$0.07 and a reduced $\chi^2$ of 1.25 (d.o.f=10). We calculated the cumulative $\chi^{2}$ of blue and black data points against the linear fitting (Figure \ref{NICERlc}, panel b) quoted above. It shows a significant increasing trend, indicating that the X-ray data during the radio active and quiet period took a different spectral evolution, perhaps suggesting they may originate from distinct emission processes. 

We propose a Crustquake-Outburst-Transient scenario to explain the results reported in this paper.
All the observed phenomena, including the frequency drifts, the emergence of radio pulsations several months 
after the onset of the X-ray outburst, and its accompanied spectral hardening in X-rays shown in Figure \ref{MJD}, could be explained in a self-consistent framework based on energy release from twisted field lines within the magnetar’s magnetosphere


\section{Radiation mechanism of the radio active period accompanied by the X-ray spectral hardening}

The origin of the magnetar radio emission is still unclear. It might arise from the open field region in the magnetosphere above the polar cap, similar to the pulsed radio emission of canonical radio pulsars (\citealt{ruderman1975,philippov2022}), or from a j-bundle, which is thicker in size and more energetic than the electron acceleration region above the polar cap (\citealt{istomin2007,beloborodov2013b}). 

The occurrence of FRB~20200428 after the onset of SGR 1935+2154’s burst forest in X-rays, its delayed radio active period, and accompanying X-ray spectral hardening behaviour shown in Figure \ref{MJD}, can be qualitatively understood within the standard scenario of magnetar outbursts (\citealt{thompson01}). A sudden energy release near the magnetar surface accompanies a crust cracking quake. Part of the energy dissipated heats up the magnetar surface and forms a trapped fireball of electron/positron pairs ($e^{\pm}$) and soft X-rays in the closed field line region of the magnetosphere near the surface. {The high outburst energy with $T\sim80$ keV (estimated from the cutoff energy in the spectrum of FRB~20200428~ (\citealt{Ioka2020})) may lead to the formation of an optically thick $e^{\pm}$ outflow surrounding the fireball, which pushes FRB photons along open field lines in the magnetosphere.} Since the fireball was formed after the crustquake powering the outburst, there should be a finite time delay between the outburst and the FRB, as observed for \sgr\/. 

The magnetic field inside a magnetar is twisted. After a crustquake, part of the magnetic energy and twist are carried over to the closed field lines. The twisting of magnetospheric field lines leads to the formation of the so-called current carrying j-bundle. Coherent radio curvature emission is generated by the charge fluctuations in the corresponding bundle. Since the magnetic field lines are twisting slowly after the crustquake, the pulsed radio emission appears with a delay compared to the outburst in X-rays. In addition, the particle return currents with Lorentz factors $\gamma\gtrsim100$ 
bombard the magnetar surface and heat it up  (\citealt{zane2019,salmi2020}). Particle bombardment onto the magnetar from the charge fluctuating bundle produces a thermally emitting hot spot on the cooling surface. This may explain the observed spectral hardening around MJD 59150. The active period of coherent radio emission depends on the time-scale of untwisting of magnetospheric field lines  (\citealt{Beloborodov2009}):
\begin{align}\label{twisteq1}
t_{\rm ev}&=\frac{BR^{2}\Delta\psi_{0}}{cV}\\ \nonumber & \simeq40\left(\frac{B}{4.4\times10^{14}\mbox{G}}\right)\left(\frac{R}{15\mbox{km}}\right)^{2}\left(\frac{\Delta\psi_{0}}{0.1}\right)\left(\frac{V}{10^{9}\mbox{volts}}\right)^{-1} \mbox{d}
\end{align}
Here $B$ is the dipolar component of the magnetic field on the polar cap, $R$ is the neutron star radius, $\Delta\psi_{0}$ is the twist angle, $c$ is the speed of light and $V$ is the voltage difference. With typical parameters of \sgr\/, Equation \ref{twisteq1} yields a radio active period of $\sim$40 days, which fits the FAST detection well. The delayed appearance of pulsed radio emission, its active period, and the accompanied X-ray spectral hardening support the scenario of crustquake-induced j-bundle current formation.

The precise radiation mechanism producing FRBs within the magnetar framework has not been fully understood yet. Two classes of models can be identified according to where the FRB is generated. The so-called ``far-away" models, according to which FRBs are generated well outside the magnetar light cylinder, rely on a relativistic outflow that drives a shock into the surrounding circumstellar medium through magnetar winds or flares, and a short duration radio burst is produced by plasma maser and synchrotron radiation processes  (\citealt{lyubarsky14,metzger19,sironi19,beloborodov20}). The predicted shock-powered X-ray fluence is close to the observed FRB~20200428  (\citealt{Margalit2020}), which indicates that part of the X-ray burst may come from shock emission. These models, however, face difficulties in explaining the double peak structure of FRB~20200428, as their narrow separation by 30 ms between the two peaks seems to be contradictory with the required larger length-scales for the synchrotron radiation from maser emission region  (\citealt{Margalit2020}). Also, the implied plasma frequency cannot be reconciled with the observed luminosity  (\citealt{lu20}).

By contrast, for models that propose FRBs generated inside the magnetar magnetospheres, several coherent radiation processes are feasible.
In the model of Ref.  (\citealt{kumar20}), FRBs are as a result of the coherent curvature radiation of $e^{\pm}$ clumps accelerated in the induced electric field, parallel to the magnetic field, at the charge starvation radius that are generated by Alfv\'{e}n waves launched into the magnetosphere after a crustquake. The coincidence of hard X-ray spikes and radio peaks can be understood within this framework. This model predicts that the FRB is produced at a distance $R_{\rm FRB}\sim20(\delta B_{10}/\lambda_{\perp4})^{6/11}R_{\rm NS}$ with $R_{\rm NS}$ being neutron star radius, $\delta B_{10}$ being the amplitude of the Alfv\'{e}n wave disturbance at the magnetar surface in units of $10^{10}$ G and $\lambda_{\perp4}$ being the wavelength perpendicular to the magnetic field in units of $10^{4}$ cm. Then, the most likely location of the double peaked FRB~20200428 is two separated open magnetic field lines at a few tens of neutron star radii from the surface. The charged clump will radiate within an emission cone of opening angle $\theta\sim\gamma^{-1}$ with $\gamma$ being the Lorentz factor gained by charged particles during acceleration along the curved field line  (\citealt{kumar2017}) and the pair separation maintains narrow spectrum  (\citealt{yang2020}). The observed downward frequency drifting with time is a natural consequence of coherent curvature radiation by clumps of pairs in the outer magnetosphere through two stream instability  (\citealt{wang2019}).  

For \sgr\/ radio pulsations' frequency drifting rate $\dot\nu_{\rm drift}\sim10$ GHz\,s$^{-1}$ and observation frequency $\nu_{\rm obs}=1.45$ GHz, Equation \ref{driftrate} yields a location $\sim4.35\times10^{9}~\mbox{cm}$, which is within the light cylinder radius $R_{\rm LC}=c/2\pi\nu=15.5\times10^{9}$ cm. Therefore, FRB~20200428 
can be 
produced at the open magnetic field region above the magnetic poles by $e^{\pm}$ pairs whereas the radio pulsations of \sgr\/ are generated by similar coherent
curvature radiation at closed field lines region in the outer magnetosphere by the formation of a j-bundle. 

Seismic energy released into the magnetosphere following crustal failure ignites FRB~20200428 right after the breaking event and the azimuthal motion of the broken platelet slowly twists the magnetic field lines, giving rise to transient pulsed radio emission through fluctuations in the magnetospheric charged particle density. For both processes a crustquake on the magnetar surface is the underlying trigger mechanism. Thus, crustquakes not only supply the right amount of energy budget powering the frequent X-ray outbursts on \sgr\/ but also lead to FRBs and transient pulsed radio emission.

\section{Conclusion}
We discover delayed radio pulsations 
associated an X-ray spectral hardening in SGR J1935+2154, i.e. it is the only confirmed radio-pulsation-related spectral change and has a time scale of months, based on the radio detections made by FAST and the X-ray detections made by NICER \& SWIFT. In this work, the observations suggest that radio emission originates from the outer magnetosphere of the magnetar, and the surface heating due to the bombardment of inward going particles from the radio emission region is responsible for the observed X-ray spectral hardening. The Crustquake-Outburst-Transient model based on observations in this work provides a unified origin story for these pulses and the first Galactic FRB.

Through a comprehensive analysis of 563 radio pulses in October 2020  and the contemporaneous X-ray behavior of the magnetar \sgr\/, we found
(i) an X-ray spectral hardening that correlates with active radio emission, while the X-ray flux faded;
(ii) narrow-band as well as frequency-drifting radio pulses, distinct from normal pulsars;
(iii) a radio emission location of in the outer magnetosphere consistent with the measured frequency drifting rate.
All of the above could be explained by the energy released from twisted field lines within the magnetar's magnetosphere, providing evidence that the magnetar radio emission mechanism differs from that of canonical radio pulsars.

\begin{acknowledgments}
This work is supported by National Natural Science Foundation of China (NSFC) Programs No. 11988101, No. 11725313, No. 11690024, No. 12273038, No. 12373051, No. 12041303, No. U1731238, No. 12173103; by CAS International Partnership Program No. 114-A11KYSB20160008; by CAS Strategic Priority Research Program No. XDB23000000; and the National Key R\&D Program of China (No. 2017YFA0402600); and the National SKA Program of China No. 2020SKA0120200.
D.L. is a New Cornerstone investigator.
P.W. acknowledges support from the National Natural Science Foundation of China (NSFC) Programs No.11988101, 12041303, the CAS Youth Interdisciplinary Team, the Youth Innovation Promotion Association CAS (id. 2021055), and the Cultivation Project for FAST Scientific Payoff and Research Achievement of CAMS-CAS.
D.F.T. acknowledges support from the Spanish grants PID2021-124581OB-I00, 2021SGR00426, CEX2020-001058-M and EU PRTR-C17.I1.
This work made use of data from FAST, a Chinese national mega-science facility built and operated by the National Astronomical Observatories, Chinese Academy of Sciences.
\end{acknowledgments}

\vspace{5mm}
\facilities{FAST, NICER, Swift}
\clearpage

\clearpage

\appendix
\section{Appendix information}
\setlength{\tabcolsep}{0.2mm}{
\renewcommand\arraystretch{1.0}

\begin{longtable}{c c c c c c c c c c c c c}%
\caption{{\bf {The properties of 563 bursts of SGR 1935+2154 measured with FAST}}}
\\
\hline%
\hline%
Pulse&\ Rot.\ period&T$_{start}$/\ T$_{obs}$&Pulse\ ToA$^{a)}$&Width&Peak\ flux&Fluence&Energy&Cntrl freq.&Bandwidth$^{b)}$&{DM$^{c)}$}\\%
$ID$& &$(MJD / hour)$&$(MJD)$&$(ms)$&$(mJy)$&$(Jy\ ms)$&$(\times 10^{27}\ erg)$&$(MHz)$&$(MHz)$&$(pc\ cm^{-3})$\\%
\hline%
\endhead%
\hline%
\endfoot%
\hline%
\endlastfoot%

1 & 1 & 59131.3843958 / 1.0 & 59131.384713925 & 4.2 (2) & 47.1 (26) & 0.020 (2) & 1.3 (1) & 1277 & 69 & 332.5 (5)\\%
2 &  &  /  & 59131.384713944 & 4.1 (4) & 24.6 (14) & 0.010 (2) & 0.6 (1) & 1236 & 86 & 332.5 (5)\\%
3 & 2 &  /  & 59131.386969843 & 4.6 (4) & 30.0 (17) & 0.014 (2) & 0.9 (1) & 1230 & 154 & 333.0 (9)\\%
4 & 3 &  /  & 59131.387270595 & 23.7 (15) & 17.3 (10) & 0.040 (4) & 2.6 (4) & 1256 & 253 & 333.3 (3)\\%
5 & 4 &  /  & 59131.388134811 & 4.9 (4) & 29.0 (16) & 0.014 (2) & 0.9 (1) &  &  & 332.2 (5)\\%
6 &  &  /  & 59131.388134844 & 14.0 (17) & 11.8 (7) & 0.016 (2) & 1.1 (2) &  &  & 332.2 (5)\\%
7 & 5 &  /  & 59131.388736181 & 3.5 (2) & 43.2 (24) & 0.015 (2) & 1.0 (1) &  &  & 332.8 (2)\\%
8 & 6 &  /  & 59131.389299918 & 4.3 (2) & 48.3 (27) & 0.021 (2) & 1.3 (1) &  &  & 332.7 (1)\\
\end{longtable}}
\begin{tablenotes}
\item[\#] $\#$ Uncertainties in parentheses refer to the last quoted digit.
\item[a)] $a)$ Arrival time of burst peak at the solar system barycenter, after correcting to the infinite frequency.
\item[b)] $b)$ A conservative 30$\%$ fractional error is assumed.
\item[c)] {$c)$ DM of the pulses obtained from the best burst alignment, calculated using the DM-Power algorithm (https://github.com/hsiuhsil/DM-power).}
\end{tablenotes}


\clearpage
\begin{longrotatetable}
\begin{deluxetable}{lcccc ccccc ccccc}
\footnotesize
\tablecaption{\bf NICER observations and fitted results
\label{table:nearbyBlazar}}
\tablehead{Obs. ID&Obs. time&Flux & Flux & Flux & Hardness ratio & T$_{BB}$& ~~~~~N$_{BB}$~~~~~ & N$_{PL}$ & Reduced $\chi^{2}$(d.o.f.)  \\
  &  & (1.5-5 keV) & (1.5-2.5 keV)& (2.5-5 keV) &\\
  & (MJD)  &  & (10$^{-12}$ erg cm$^{-2}$ s$^{-1}$)&  &  & (keV) & (10$^{-5}$)  & (10$^{-4}$) &}
\startdata
\hline
3020560102&\multirow{4}{*}{58969.06$\pm$0.48}&\multirow{4}{*}{4.22$\pm$0.07}&\multirow{4}{*}{2.12$\pm$0.04}&\multirow{4}{*}{2.10$\pm$0.05}&\multirow{4}{*}{0.99$\pm$0.03}&\multirow{4}{*}{0.51$\pm$0.01}&\multirow{4}{*}{5.66$\pm$0.28}&\multirow{4}{*}{2.94$\pm$0.48}&\multirow{4}{*}{1.12 (203)}\\
3655010101\\
3655010102\\
3020560103\\
\hline
3020560104&\multirow{4}{*}{58985.10$\pm$4.49}&\multirow{4}{*}{3.06$\pm$0.06}&\multirow{4}{*}{1.60$\pm$0.03}&\multirow{4}{*}{1.46$\pm$0.03}&\multirow{4}{*}{0.92$\pm$0.03}&\multirow{4}{*}{0.41$\pm$0.01}&\multirow{4}{*}{4.03$\pm$0.15}&\multirow{4}{*}{3.36$\pm$0.28}&\multirow{4}{*}{1.07 (201)}\\
3655010201\\
3020560105\\
3020560106\\
\hline
3020560107&\multirow{1}{*}{58991.04$\pm$0.03}&\multirow{1}{*}{2.14$\pm$0.07}&\multirow{1}{*}{1.21$\pm$0.03}&\multirow{1}{*}{0.93$\pm$0.05}&\multirow{1}{*}{0.77$\pm$0.04}&0.46$\pm$0.02&3.45$\pm$0.25&1.26$\pm$0.5&0.94 (138)\\
\hline
3020560108&\multirow{1}{*}{58992.01$\pm$0.02}&\multirow{1}{*}{3.12$\pm$0.10}&\multirow{1}{*}{1.49$\pm$0.05}&\multirow{1}{*}{1.63$\pm$0.06}&\multirow{1}{*}{1.09$\pm$0.05}&0.39$\pm$0.03&3.29$\pm$0.29&4.33$\pm$0.6&0.99 (170)\\
\hline
3020560109&\multirow{5}{*}{58997.34$\pm$2.74}&\multirow{5}{*}{2.02$\pm$0.06}&\multirow{5}{*}{1.19$\pm$0.03}&\multirow{5}{*}{0.83$\pm$0.04}&\multirow{5}{*}{0.70$\pm$0.04}&\multirow{5}{*}{0.47$\pm$0.02}&\multirow{5}{*}{3.64$\pm$0.2}&\multirow{5}{*}{0.68$\pm$0.37}&\multirow{5}{*}{0.97 (142)}\\
3020560110\\
3020560111\\
3020560112\\
3020560113\\
\hline
3020560114&\multirow{1}{*}{59001.11$\pm$0.02}&\multirow{1}{*}{2.09$\pm$0.09}&\multirow{1}{*}{1.13$\pm$0.04}&\multirow{1}{*}{0.95$\pm$0.06}&\multirow{1}{*}{0.84$\pm$0.06}&0.43$\pm$0.03&3.01$\pm$0.28&1.82$\pm$0.55&1.07 (74)\\
\hline
3020560115&\multirow{5}{*}{59004.60$\pm$1.61}&\multirow{5}{*}{1.47$\pm$0.06}&\multirow{5}{*}{0.91$\pm$0.03}&\multirow{5}{*}{0.56$\pm$0.04}&\multirow{5}{*}{0.62$\pm$0.05}&\multirow{5}{*}{0.49$\pm$0.01}&\multirow{5}{*}{3.01$\pm$0.1}&\multirow{5}{*}{-}&\multirow{5}{*}{1.37 (105)}\\
3020560116\\
3020560117\\
3020560118\\
3020560119\\
\hline
3655010301&\multirow{2}{*}{59018.13$\pm$0.23}&\multirow{2}{*}{1.48$\pm$0.05}&\multirow{2}{*}{0.95$\pm$0.03}&\multirow{2}{*}{0.53$\pm$0.04}&\multirow{2}{*}{0.56$\pm$0.04}&\multirow{2}{*}{0.47$\pm$0.01}&\multirow{2}{*}{3.17$\pm$0.09}&\multirow{2}{*}{-}&\multirow{2}{*}{0.95 (214)}\\
3655010302\\
\hline
3655010303&\multirow{4}{*}{59020.72$\pm$1.71}&\multirow{4}{*}{1.51$\pm$0.05}&\multirow{4}{*}{1.01$\pm$0.03}&\multirow{4}{*}{0.51$\pm$0.03}&\multirow{4}{*}{0.51$\pm$0.03}&\multirow{4}{*}{0.45$\pm$0.01}&\multirow{4}{*}{3.37$\pm$0.13}&\multirow{4}{*}{0.05$\pm$0.17}&\multirow{4}{*}{0.92 (161)}\\
3020560120\\
3020560121\\
3020560122\\
\hline
3020560123&\multirow{3}{*}{59024.50$\pm$1.42}&\multirow{3}{*}{1.43$\pm$0.05}&\multirow{3}{*}{0.94$\pm$0.03}&\multirow{3}{*}{0.50$\pm$0.03}&\multirow{3}{*}{0.53$\pm$0.04}&\multirow{3}{*}{0.46$\pm$0.01}&\multirow{3}{*}{3.13$\pm$0.09}&\multirow{3}{*}{-}&\multirow{3}{*}{1.21 (125)}\\
3020560124\\
3020560125\\
\hline
3020560126&\multirow{4}{*}{59029.14$\pm$1.54}&\multirow{4}{*}{1.27$\pm$0.05}&\multirow{4}{*}{0.84$\pm$0.03}&\multirow{4}{*}{0.43$\pm$0.03}&\multirow{4}{*}{0.51$\pm$0.04}&\multirow{4}{*}{0.46$\pm$0.01}&\multirow{4}{*}{2.83$\pm$0.1}&\multirow{4}{*}{-}&\multirow{4}{*}{1.37 (137)}\\
3020560127\\
3020560128\\
3020560129\\
\hline
3020560130&\multirow{7}{*}{59037.18$\pm$5.66}&\multirow{7}{*}{1.47$\pm$0.05}&\multirow{7}{*}{0.99$\pm$0.03}&\multirow{7}{*}{0.48$\pm$0.03}&\multirow{7}{*}{0.49$\pm$0.04}&\multirow{7}{*}{0.44$\pm$0.02}&\multirow{7}{*}{3.32$\pm$0.14}&\multirow{7}{*}{0.08$\pm$0.18}&\multirow{7}{*}{1.03 (129)}\\
3020560131\\
3020560132\\
3020560133\\
3020560134\\
3020560135\\
3020560136\\
3020560137\\
\hline
3020560138&\multirow{4}{*}{59047.34$\pm$1.68}&\multirow{4}{*}{1.11$\pm$0.05}&\multirow{4}{*}{0.79$\pm$0.03}&\multirow{4}{*}{0.32$\pm$0.03}&\multirow{4}{*}{0.41$\pm$0.04}&\multirow{4}{*}{0.42$\pm$0.01}&\multirow{4}{*}{2.76$\pm$0.06}&\multirow{4}{*}{-}&\multirow{4}{*}{1.34 (134)}\\
3020560139\\
3020560140\\
3020560141\\
\hline
3020560142&\multirow{4}{*}{59059.64$\pm$3.58}&\multirow{4}{*}{1.24$\pm$0.07}&\multirow{4}{*}{0.86$\pm$0.04}&\multirow{4}{*}{0.38$\pm$0.04}&\multirow{4}{*}{0.44$\pm$0.06}&\multirow{4}{*}{0.43$\pm$0.02}&\multirow{4}{*}{2.97$\pm$0.15}&\multirow{4}{*}{-}&\multirow{4}{*}{0.93 (100)}\\
3020560143\\
3020560144\\
3020560145\\
\hline
3020560146&\multirow{3}{*}{59065.97$\pm$1.57}&\multirow{3}{*}{1.89$\pm$0.06}&\multirow{3}{*}{1.00$\pm$0.03}&\multirow{3}{*}{0.89$\pm$0.05}&\multirow{3}{*}{0.90$\pm$0.05}&\multirow{3}{*}{0.47$\pm$0.03}&\multirow{3}{*}{2.63$\pm$0.22}&\multirow{3}{*}{1.46$\pm$0.42}&\multirow{3}{*}{0.93 (121)}\\
3020560147\\
3020560148\\
\hline
3020560149&\multirow{5}{*}{59103.56$\pm$13.66}&\multirow{5}{*}{1.80$\pm$0.06}&\multirow{5}{*}{1.02$\pm$0.03}&\multirow{5}{*}{0.79$\pm$0.03}&\multirow{5}{*}{0.77$\pm$0.04}&\multirow{5}{*}{0.42$\pm$0.02}&\multirow{5}{*}{2.84$\pm$0.17}&\multirow{5}{*}{1.49$\pm$0.31}&\multirow{5}{*}{1.13 (109)}\\
3020560150\\
3020560151\\
3020560152\\
3020560153\\
\hline
3655010401&\multirow{1}{*}{59128.06$\pm$0.05}&\multirow{1}{*}{2.16$\pm$0.05}&\multirow{1}{*}{1.10$\pm$0.02}&\multirow{1}{*}{1.06$\pm$0.03}&\multirow{1}{*}{0.96$\pm$0.03}&0.42$\pm$0.02&2.68$\pm$0.14&2.4$\pm$0.29&0.99 (209)\\
\hline
3655010402&\multirow{1}{*}{59129.02$\pm$0.08}&\multirow{1}{*}{1.31$\pm$0.03}&\multirow{1}{*}{0.88$\pm$0.02}&\multirow{1}{*}{0.43$\pm$0.02}&\multirow{1}{*}{0.49$\pm$0.03}&0.43$\pm$0.02&2.94$\pm$0.16&0.15$\pm$0.31&0.80 (166)\\
\hline
3020560154&\multirow{6}{*}{59135.22$\pm$3.68}&\multirow{6}{*}{1.42$\pm$0.04}&\multirow{6}{*}{0.98$\pm$0.03}&\multirow{6}{*}{0.44$\pm$0.03}&\multirow{6}{*}{0.45$\pm$0.03}&\multirow{6}{*}{0.43$\pm$0.01}&\multirow{6}{*}{3.35$\pm$0.11}&\multirow{6}{*}{-}&\multirow{6}{*}{1.05 (128)}\\
3020560155\\
3020560156\\
3020560157\\
3020560158\\
3020560159\\
\hline
3020560160&\multirow{1}{*}{59139.03$\pm$0.06}&\multirow{1}{*}{1.44$\pm$0.04}&\multirow{1}{*}{0.96$\pm$0.02}&\multirow{1}{*}{0.48$\pm$0.03}&\multirow{1}{*}{0.50$\pm$0.03}&\multirow{1}{*}{0.45$\pm$0.01}&\multirow{1}{*}{3.24$\pm$0.09}&\multirow{1}{*}{-}&\multirow{1}{*}{0.99 (166)}\\
\hline
3020560161&\multirow{1}{*}{59140.76$\pm$0.04}&\multirow{1}{*}{1.05$\pm$0.05}&\multirow{1}{*}{0.77$\pm$0.03}&\multirow{1}{*}{0.27$\pm$0.03}&\multirow{1}{*}{0.35$\pm$0.05}&\multirow{1}{*}{0.4$\pm$0.02}&\multirow{1}{*}{2.82$\pm$0.14}&\multirow{1}{*}{-}&\multirow{1}{*}{1.08 (87)}\\
\hline
3020560162&\multirow{1}{*}{59141.02$\pm$0.08}&\multirow{1}{*}{1.55$\pm$0.04}&\multirow{1}{*}{0.95$\pm$0.02}&\multirow{1}{*}{0.60$\pm$0.02}&\multirow{1}{*}{0.63$\pm$0.03}&\multirow{1}{*}{0.43$\pm$0.02}&\multirow{1}{*}{2.93$\pm$0.15}&\multirow{1}{*}{0.74$\pm$0.34}&\multirow{1}{*}{0.89 (189)}\\
\hline
3020560163&\multirow{1}{*}{59141.99$\pm$0.10}&\multirow{1}{*}{1.33$\pm$0.04}&\multirow{1}{*}{0.91$\pm$0.02}&\multirow{1}{*}{0.42$\pm$0.03}&\multirow{1}{*}{0.46$\pm$0.03}&\multirow{1}{*}{0.44$\pm$0.01}&\multirow{1}{*}{3.09$\pm$0.11}&\multirow{1}{*}{-}&\multirow{1}{*}{1.00 (155)}\\
\hline
3020560164&\multirow{1}{*}{59143.02$\pm$0.04}&\multirow{1}{*}{1.56$\pm$0.05}&\multirow{1}{*}{0.95$\pm$0.03}&\multirow{1}{*}{0.62$\pm$0.04}&\multirow{1}{*}{0.65$\pm$0.04}&\multirow{1}{*}{0.45$\pm$0.03}&\multirow{1}{*}{2.91$\pm$0.24}&\multirow{1}{*}{0.61$\pm$0.54}&\multirow{1}{*}{0.87 (111)}\\
\hline
3020560165&\multirow{1}{*}{59143.99$\pm$0.03}&\multirow{1}{*}{1.60$\pm$0.06}&\multirow{1}{*}{0.98$\pm$0.03}&\multirow{1}{*}{0.61$\pm$0.04}&\multirow{1}{*}{0.62$\pm$0.05}&\multirow{1}{*}{0.45$\pm$0.02}&\multirow{1}{*}{3.08$\pm$0.19}&\multirow{1}{*}{0.46$\pm$0.37}&\multirow{1}{*}{1.10 (90)}\\
\hline
3020560166&\multirow{1}{*}{59146.57$\pm$0.04}&\multirow{1}{*}{2.31$\pm$0.05}&\multirow{1}{*}{1.14$\pm$0.03}&\multirow{1}{*}{1.17$\pm$0.03}&\multirow{1}{*}{1.03$\pm$0.04}&\multirow{1}{*}{0.41$\pm$0.02}&\multirow{1}{*}{2.61$\pm$0.15}&\multirow{1}{*}{2.83$\pm$0.29}&\multirow{1}{*}{1.09 (172)}\\
\hline
3020560167&\multirow{1}{*}{59147.47$\pm$0.03}&\multirow{1}{*}{2.56$\pm$0.06}&\multirow{1}{*}{1.12$\pm$0.03}&\multirow{1}{*}{1.44$\pm$0.04}&\multirow{1}{*}{1.29$\pm$0.04}&\multirow{1}{*}{0.38$\pm$0.03}&\multirow{1}{*}{2.04$\pm$0.17}&\multirow{1}{*}{4.1$\pm$0.3}&\multirow{1}{*}{1.06 (147)}\\
\hline
3020560168&\multirow{1}{*}{59147.99$\pm$0.05}&\multirow{1}{*}{2.32$\pm$0.05}&\multirow{1}{*}{1.07$\pm$0.02}&\multirow{1}{*}{1.25$\pm$0.03}&\multirow{1}{*}{1.16$\pm$0.04}&\multirow{1}{*}{0.38$\pm$0.02}&\multirow{1}{*}{2.24$\pm$0.15}&\multirow{1}{*}{3.43$\pm$0.27}&\multirow{1}{*}{0.99 (183)}\\
\hline
3020560169&\multirow{1}{*}{59150.51$\pm$0.01}&\multirow{1}{*}{2.30$\pm$0.09}&\multirow{1}{*}{1.18$\pm$0.05}&\multirow{1}{*}{1.12$\pm$0.06}&\multirow{1}{*}{0.95$\pm$0.07}&\multirow{1}{*}{0.47$\pm$0.03}&\multirow{1}{*}{3.02$\pm$0.32}&\multirow{1}{*}{1.95$\pm$0.58}&\multirow{1}{*}{1.10 (56)}\\
\hline
3020560170&\multirow{1}{*}{59165.63$\pm$0.02}&\multirow{1}{*}{1.59$\pm$0.07}&\multirow{1}{*}{1.00$\pm$0.04}&\multirow{1}{*}{0.60$\pm$0.05}&\multirow{1}{*}{0.60$\pm$0.05}&\multirow{1}{*}{0.45$\pm$0.02}&\multirow{1}{*}{3.16$\pm$0.22}&\multirow{1}{*}{0.42$\pm$0.38}&\multirow{1}{*}{1.08 (48)}\\
\hline
3020560171&\multirow{1}{*}{59166.53$\pm$0.04}&\multirow{1}{*}{1.38$\pm$0.04}&\multirow{1}{*}{0.93$\pm$0.03}&\multirow{1}{*}{0.45$\pm$0.03}&\multirow{1}{*}{0.49$\pm$0.03}&\multirow{1}{*}{0.45$\pm$0.01}&\multirow{1}{*}{3.14$\pm$0.09}&\multirow{1}{*}{-}&\multirow{1}{*}{0.91 (113)}\\
\hline
3020560172&\multirow{1}{*}{59172.47$\pm$0.03}&\multirow{1}{*}{1.33$\pm$0.05}&\multirow{1}{*}{0.87$\pm$0.03}&\multirow{1}{*}{0.45$\pm$0.03}&\multirow{1}{*}{0.52$\pm$0.04}&\multirow{1}{*}{0.42$\pm$0.02}&\multirow{1}{*}{2.87$\pm$0.16}&\multirow{1}{*}{0.34$\pm$0.29}&\multirow{1}{*}{0.80 (78)}\\
\hline
3020560173&\multirow{1}{*}{59173.50$\pm$0.08}&\multirow{1}{*}{1.81$\pm$0.04}&\multirow{1}{*}{1.03$\pm$0.02}&\multirow{1}{*}{0.78$\pm$0.02}&\multirow{1}{*}{0.76$\pm$0.03}&\multirow{1}{*}{0.43$\pm$0.01}&\multirow{1}{*}{2.9$\pm$0.11}&\multirow{1}{*}{1.36$\pm$0.2}&\multirow{1}{*}{1.04 (225)}\\
\enddata
\label{table:nicer}
\end{deluxetable}
\begin{flushleft}
\footnotesize 
{\bf Notes.}\\ 
(1): ‘-’ means parameter could not be well constrained.\\
\end{flushleft}
\end{longrotatetable}

\clearpage
\begin{longrotatetable}
\begin{deluxetable}{lcccc ccccc ccccc}
\footnotesize
\tablecaption{\bf SWIFT-XRT observations and fitted results}
\tablehead{Obs. ID&Obs. time&Flux & Flux & Flux & Hardness ratio & T$_{BB}$ &~~~~~ N$_{BB}$ ~~~~~& N$_{PL}$ & Reduced $\chi^{2}$(d.o.f.)  \\
  &  & (1.5-5 keV) & (1.5-2.5 keV)& (2.5-5 keV) &\\
  & (MJD)  &  & (10$^{-12}$ erg cm$^{-2}$ s$^{-1}$)&  &  & (keV) & (10$^{-5}$)  & (10$^{-4}$) &}
\startdata
00033349046&\multirow{9}{*}{58970.75$\pm$2.19}&\multirow{9}{*}{3.33$\pm$0.14}&\multirow{9}{*}{1.61$\pm$0.07}&\multirow{9}{*}{1.72$\pm$0.09}&\multirow{9}{*}{1.07$\pm$0.07}&\multirow{9}{*}{0.45$\pm$0.03}&\multirow{9}{*}{3.60$\pm$0.35}&\multirow{9}{*}{3.88$\pm$0.43}&\multirow{9}{*}{1.43 (39)}
\\
00033349047\\
00033349048\\
00033349049\\
00033349050\\
00033349051\\
00033349053\\
00033349052\\
00033349054\\
\hline
00033349056&\multirow{5}{*}{58976.15$\pm$3.05}&\multirow{5}{*}{2.57$\pm$0.12}&\multirow{5}{*}{1.32$\pm$0.07}&\multirow{5}{*}{1.25$\pm$0.08}&\multirow{5}{*}{0.95$\pm$0.08}&\multirow{5}{*}{0.50$\pm$0.03}&\multirow{5}{*}{3.54$\pm$0.29}&\multirow{5}{*}{1.79$\pm$0.33}&\multirow{5}{*}{0.56 (18)}
\\
00033349058\\
00033349060\\
00033349061\\
00033349062\\
\hline
00033349067&\multirow{8}{*}{59018.51$\pm$20.99}&\multirow{8}{*}{1.66$\pm$0.08}&\multirow{8}{*}{0.95$\pm$0.05}&\multirow{8}{*}{0.71$\pm$0.05}&\multirow{8}{*}{0.74$\pm$0.06}&\multirow{8}{*}{0.45$\pm$0.03}&\multirow{8}{*}{2.75$\pm$0.20}&\multirow{8}{*}{1.02$\pm$0.23}&\multirow{8}{*}{1.00 (18)}
\\
00033349068\\
00033349069\\
00089040001\\
00033349070\\
00033349071\\
00033349072\\
00033349073\\
\hline
00033349074&\multirow{9}{*}{59071.79$\pm$25.26}&\multirow{9}{*}{1.55$\pm$0.08}&\multirow{9}{*}{0.92$\pm$0.05}&\multirow{9}{*}{0.63$\pm$0.04}&\multirow{9}{*}{0.68$\pm$0.06}&\multirow{9}{*}{0.39$\pm$0.03}&\multirow{9}{*}{2.80$\pm$0.30}&\multirow{9}{*}{1.25$\pm$0.27}&\multirow{9}{*}{0.64 (13)}
\\
00033349075\\
00033349077\\
00033349078\\
00033349079\\
00033349080\\
00033349081\\
00033349082\\
00033349083\\
\hline
00033349089&\multirow{4}{*}{59133.62$\pm$1.36}&\multirow{4}{*}{1.40$\pm$0.14}&\multirow{4}{*}{0.88$\pm$0.09}&\multirow{4}{*}{0.52$\pm$0.09}&\multirow{4}{*}{0.60$\pm$0.12}&\multirow{4}{*}{0.48$\pm$0.04}&\multirow{4}{*}{2.91$\pm$0.38}&\multirow{4}{*}{0.00$\pm$0.37}&\multirow{4}{*}{0.41 (2)}
\\
00033349093\\
00033349094\\
00033349095\\
\enddata
\label{table:swift}
\end{deluxetable}
\begin{flushleft}
\footnotesize 
{\bf Notes.}\\ 
(1): ‘-’ means parameter could not be well constrained.\\
\end{flushleft}
\end{longrotatetable}


\bibliography{sample631}{}
\bibliographystyle{aasjournal}

\end{document}